# Covid-19 pandemic and the unprecedented mobilisation of scholarly efforts prompted by a health crisis: Scientometric comparisons across SARS, MERS and 2019-nCov literature

Milad Haghani, Michiel C. J. Bliemer

*The University of Sydney Business School, NSW, Australia*

**Abstract**

During the current century, each major coronavirus outbreak has triggered a quick and immediate surge of academic publications on this topic. The spike in research publications following the 2019 Novel Coronavirus (Covid-19) outbreak, however, has been like no other. The global crisis caused by the Covid-19 pandemic has mobilised scientific efforts in an unprecedented way. In less than five months, more than 12,000 research items have been indexed while the number increasing every day. With the crisis affecting all aspects of life, research on Covid-19 seems to have become a focal point of interest across many academic disciplines. Here, scientometric aspects of the Covid-19 literature are analysed and contrasted with those of the two previous major Coronavirus diseases, i.e. Severe Acute Respiratory Syndrome (SARS) and Middle East Respiratory Syndrome (MERS). The focus is on the co-occurrence of key-terms, bibliographic coupling and citation relations of journals and collaborations between countries. Certain recurring patterns across all three literatures were discovered. All three outbreaks have commonly generated three distinct and major cohort of studies: (i) studies linked to the public health response and epidemic control, (ii) studies associated with the chemical constitution of the virus and (iii) studies related to treatment, vaccine and clinical care. While studies affiliated with the category (i) seem to have been the first to emerge, they overall received least numbers of citations compared to those of the two other categories. Covid-19 studies seem to have been distributed across a broader variety of journals and subject areas. Clear links are observed between the geographical origins of each outbreak or the local geographical severity of each outbreak and the magnitude of research originated from regions. Covid-19 studies also display the involvement of authors from a broader variety of countries compared to SARS and MRS.

## 1. Introduction

On December 31, 2019 an official case of a novel respiratory diseases of the category of Coronaviruses, named Covid-19, was reported in Wuhan, China, marking the beginning of what proved to be one of the direst and most devastating viral outbreaks in the modern history (Sohrabi *et al.* 2020, Wang *et al.* 2020). This was immediately followed by an unprecedented and swift response of the academic community to address various dimensions of this health crisis and prompted an avalanche of scholarly publications on this topic (Golinelli *et al.* 2020, Haghani *et al.* 2020, Kagan *et al.* 2020). In less than five months, more than 12,000 publications on this topic have already been indexed by Scopus with the number increasing figuratively every day in considerable increments (Torres-Salinas *et al.* 2020). And this figure does not even include many more publications available in various repositories, including CORD-19 (Colavizza *et al.* 2020), in the form of preprints awaiting peer review by their respective journals. Such explosion of research on a single topic and the off-the-charts surge in the rate of publications are arguably unprecedented trends in the history of scholarly publications. An article published by Science on May 13, 2020, referred to this phenomenon as one that is "among the biggest explosions of scientific literature ever" (Brainard 2020). It highlighted how, in the face of this phenomenon, it has become extremely challenging for scientists to stay abreast of the latest developments. This has made the importance of research synthesis more tangible than ever and has even resulted in the development of several computational research mining tools for this very topic utilising methods such as Artificial Intelligence (AI). Among such efforts is a research synthesis powered by AI algorithms which has harvested datapoints from a large number the CORD-19 articles and categorised them (Brainard 2020).

Though the impact of the Covid-19 health crisis has marked it as a rather unique milestone in the history disease outbreaks, the world, prior to this, was not a stranger with Coronavirus disease outbreaks (McIntosh 1974, Myint 1994, Cavanagh 2005, Lim *et al.* 2016, Chen *et al.* 2020). Prior to 2020, two major outbreaks of this family of viruses had already been reported with at least one of them carrying the official label of a "global pandemic" (Wang *et al.* 2020). On November 16, 2002, the first case of the Severe Acute Respiratory Syndrome (SARS) disease was reported in the Guangdong province in southern China, which by 2003, swiftly spread from continent to continent, prompting the World Health Organisation to declare it as a pandemic. In fact, SARS is known to be "the First Pandemic of the 21st Century" (Cherry and Krogstad 2004). Nearly ten years later, on June 13, 2012, the first case of the Middle Eastern Respiratory Syndrome (MERS) disease was discovered in Jeddah, Saudi Arabia. These two constituted the two most major Coronavirus outbreaks until Covid-19 came along. Similar to Covid-19, though at a much smaller scale, each of these previous outbreaks generated a literature of their own (Kostoff and Morse 2011).

In the face of the flood of scholarly outputs on Covid-19, and along with the conventional review and research synthesis studies (Chang *et al.* 2020, Chen *et al.* 2020, Cortegiani *et al.* 2020), scientometric (Colavizza *et al.* 2020) and bibliometric methods (Bonilla-Aldana *et al.* 2020, Hossain 2020) have also gained traction in documenting and analysing the rapid developments of this literature (Chahrour *et al.* 2020, Dehghanbanadaki *et al.* 2020, Haghani *et al.* 2020, Kumar 2020, Le Bras *et al.* 2020). Here in this work, the literatures of these three major Coronavirus diseases are disentangled and analysed in a comparative way and from scientometric perspectives. The aim is to discover possible similarities and discrepancies across these three segments of the Coronavirus literature, and to discover whether there are recurring patterns in terms of magnitude, temporal evolution and the shape of these three literatures that were each developed in response to a disease outbreak. The main focus of the analyses is on keyword co-occurrences, bibliographic coupling and citation relations of sources and collaborations between countries.

## 2. Methods, data and general statistics

To compare the scientometric aspects of the studies on SARS, MERS and Covid-19, three separate datasets of publications on these three topics were retrieved from Scopus through three separate search strategies. The decision on which general database to use (e.g. Web of Science (WoS) or Scopus) was mainly made based on the number of indexed Covid-19 studies as the sector of the literature that is currently emerging (compared to the literatures on SARS and MERS that are better established). At the time of the data retrieval, WoS had indexed slightly less than 5,000 research items on Covid-19, while the number of items in Scopus neared 12,000. Given the fact that the Scopus database was considerably more up to date in that area, this database was set as the main source of data extraction in this work. Therefore, for the sake of consistency, the data for SARS and MERS were also extracted from Scopus.

The search strategies were devised in a way to minimise the possible overlap between the datasets on SARS, MERS and Covid-19 and to disentangle the three datasets to the most possible extent. Preliminary inspection of the literature on each three topics determined a set of distinct keywords that would return the target literature with reasonable specificity and sensitivity. In each search, key terms associated with the other literatures were combined with the logical operator “AND NOT” in order to avoid the overlap. The lower bound of the time span for each search was set with consideration of the year where the first outbreak of each virus took place. The query string associated with each dataset are as follows:

**SARS**: *( TITLE-ABS-KEY ( ( ( "Severe acute respiratory syndrome" OR "SARS" ) AND ( coronavirus* ) ) OR ( "SARS virus" OR "SARS disease" OR "Severe acute respiratory syndrome disease" OR "Severe acute respiratory syndrome virus" OR "SARS-Cov") ) AND NOT TITLE-ABS-KEY ( ( "covid" OR "nCov" OR "Covid-19" OR "covid19" OR "SARS-Cov-2" OR "Severe acute respiratory syndrome-2" OR "MERS" OR "middle east respiratory syndrome" ) ) ) AND PUBYEAR > 2001*

**MERS**: *( TITLE-ABS-KEY ( ( ( "Middle east respiratory syndrome" OR "MERS" ) AND ( coronavirus ) ) OR ( "MERS-Cov" OR "MERS virus" OR "MERS disease" OR "Middle east respiratory syndrome virus" OR "Middle east respiratory syndrome disease" ) ) AND NOT TITLE-ABS-KEY ( ( "nCov" OR "Covid-19" OR "covid19" OR "SARS-Cov" OR "SARS-Cov-2" OR "SARS" OR "Severe acute respiratory syndrome" ) ) ) AND PUBYEAR > 2011*

**Covid-19**: *TITLE-ABS-KEY ( "covid-19" OR "covid19" OR "coronavirus disease 2019" OR "2019-nCov" OR "Novel Coronavirus" OR "Novel Corona virus" OR "SARS-Cov-2" ) AND PUBYEAR > 2018*

The search was last time updated on 24 May 2020 where it returned 5,907 items on SARS, 1,752 items on MERS and 11,859 items on Covid-19. Figures 1, 2 and 3 show the distribution of the studies on SARS, MERS and Covid-19, respectively, across subject areas. Figure 4(a) also shows the composition of the Covid-19 literature in terms of the document types, demonstrating that only nearly 50% of the studies on this topic have so far been in the form of full-length articles, while letters, notes, reviews, and other document formats constitute a large portion (i.e. nearly half) of the literature on this topic at the time of this investigation.

Full records of the three datasets on SARS, MERS and Covid-19 were retrieved in *CSV Excel* format from Scopus, all on the same day. This included the citation information, bibliographic information, abstract and keywords, funding details and the references. The Scopus restriction of maximum 2,000 document to export posed challenges for the retrieval of the SARS and Covid-19 datasets whose size were bigger than 2,000 documents. For the SARS dataset, the challenge was circumvented by further limiting the search to specific year(s), in separate bundles, in a way that the size of each bundle was less than 2,000 items, therefore allowing us to export the items of each bundle separately. The extraction of the Covid-19 dataset, however, posed a further layer of complication, given that nearly all studies of Covid-19 have been published in one year, i.e. 2020. Therefore, the year of publication could not be used as a criterion to form a set of mutually exclusive smaller-size exportable bundles for this literature. To decompose the search outcome to bundles of 2,000 documents or less, the following strategy was adopted. The *Document Type* was used to initially limit the search to mutually exclusive (non-overlapping) categories. First, the search was limited to "*Review* or *Short Survey* or *Erratum* or *Conference Paper* or *Data paper*". This formed a set of 1,267 documents which was extracted in one single export (see Figure 4(a) for details of the number of items within each *Document Type* category). Subsequently, the search was set back to original and was limited to *Notes* (1,067 items) and then to *Editorial* (1,270 items). With each set of these two subsets being smaller than 2,000, they were exported separately. There were 2,564 documents of *Letter* type. This set was further decomposed to two mutually exclusive subsets based on the *Publication Stage* criterion (1,539 *Article in Press*, and 1,025 *Final*) and was retrieved in two separate exports. For the remaining 5,691 *Article* documents, the following strategy was devised. Of the 5,691 items, 2,944 were *Article in Press* and 2,747 were *Final*. First, the 2,944 *Article in Press* items were considered. The list of those studies was sorted as *First Author (A-Z)* and the first 2,000 items were extracted in one export. Then the list was sorted as *First Author (Z-A)* and the first 944 items were exported. Similar strategy was utilised to extract the remaining 2,747 *Final* documents.

A supplementary search was also conducted on the general topic of coronaviruses using the string *TITLE-ABS-KEY ( "Coronavirus*" ) AND PUBYEAR > 1985* which returned 24,620 documents on the same day. Only the data related to the number of documents by year was extracted for this search.

Figure 4(b) shows the temporal distribution of the studies on the general topic of Coronaviruses. The graph clearly shows spikes of publication coinciding with the years when SARS, MERS and Covid-19 outbreaks took place. The first spike is related to the SARS outbreak in 2002 which is reflected in an immediate and substantial increase in the number of publications on Coronaviruses from 2002 to 2003. The increase continued, though at a slower rate, to 2004 and was then followed by a gradual decline till 2012. The 2012 MERS outbreak triggered another spike in the number of publications on Coronaviruses, though not as large as that of the SARS. The intensification of attention to this topic this time lasted for about three years till 2015 before another decline began. The spike of Coronavirus studies prompted by the Covid-19 outbreak, however, seem to have been occurring at a completely different scale which can be deemed unprecedented in the history of Coronavirus studies. The number of studies emerged in the first five months of 2020 nears an equivalent of the 70% of the total sizer of Coronavirus literature during more than 50 years (1968-2019). In Figure 4(c), the temporal distribution of the SARS, MERS and Covid-19 studies have been shown separately according to the three datasets explained earlier. Note that, the quantities associated with SARS and MERS are represented by the left vertical axis whereas that of the Covid-19 is represented by the right vertical axis, with a scale ten times bigger than the scale of the left axis.

The history of previous Coronavirus research has suggested that the number of studies will likely keep rising for at least a few years before it peaks. But given the unprecedented magnitude of research and the explosive rate of publications since the begging of 2020, it would be interesting to observe whether this pattern would repeat itself and whether the peak would occur at an earlier or later stage compared to those of the previous outbreaks, a question whose answer will only be determined by time.

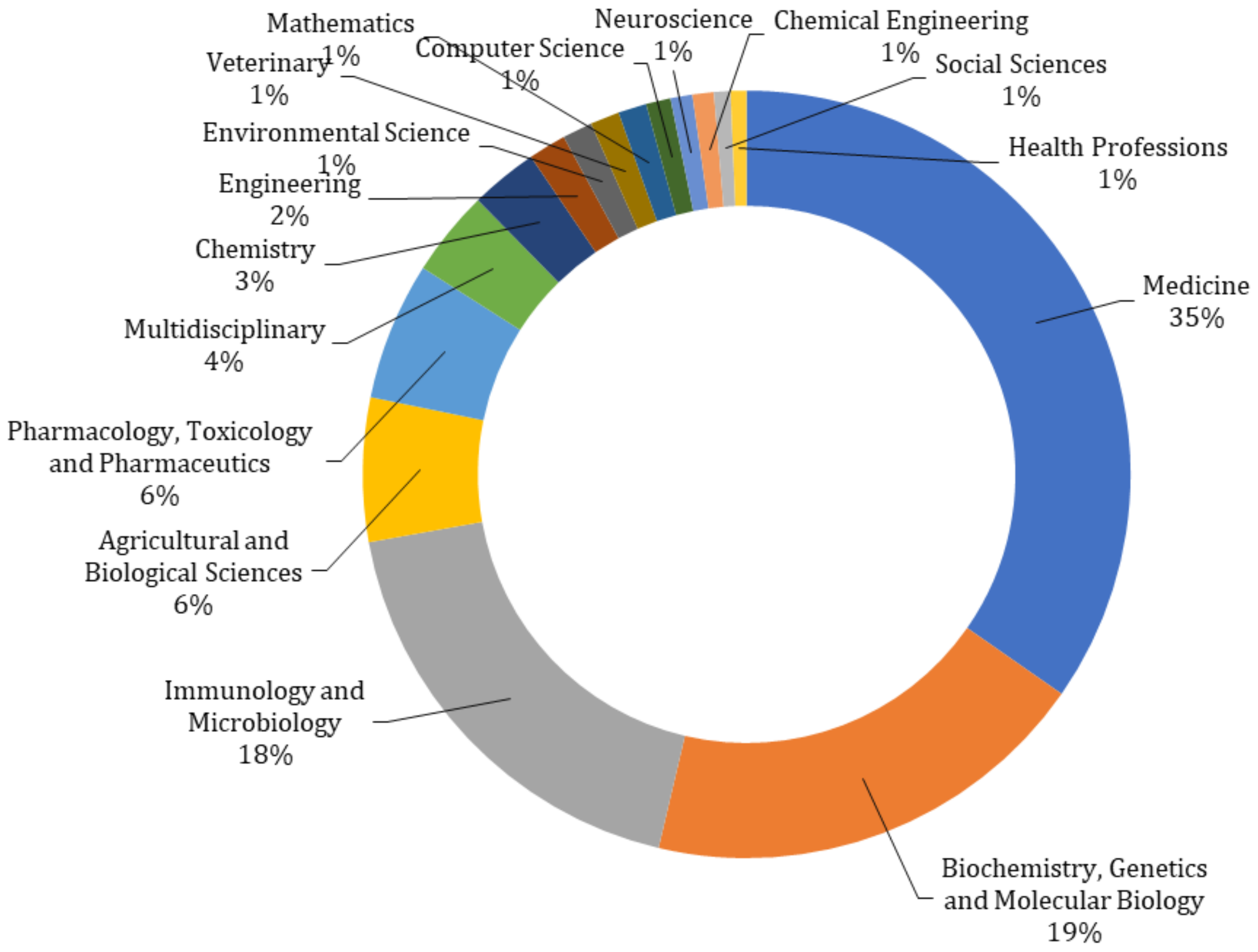

**Figure 1** Distribution of SARS studies across *Subject Areas*.

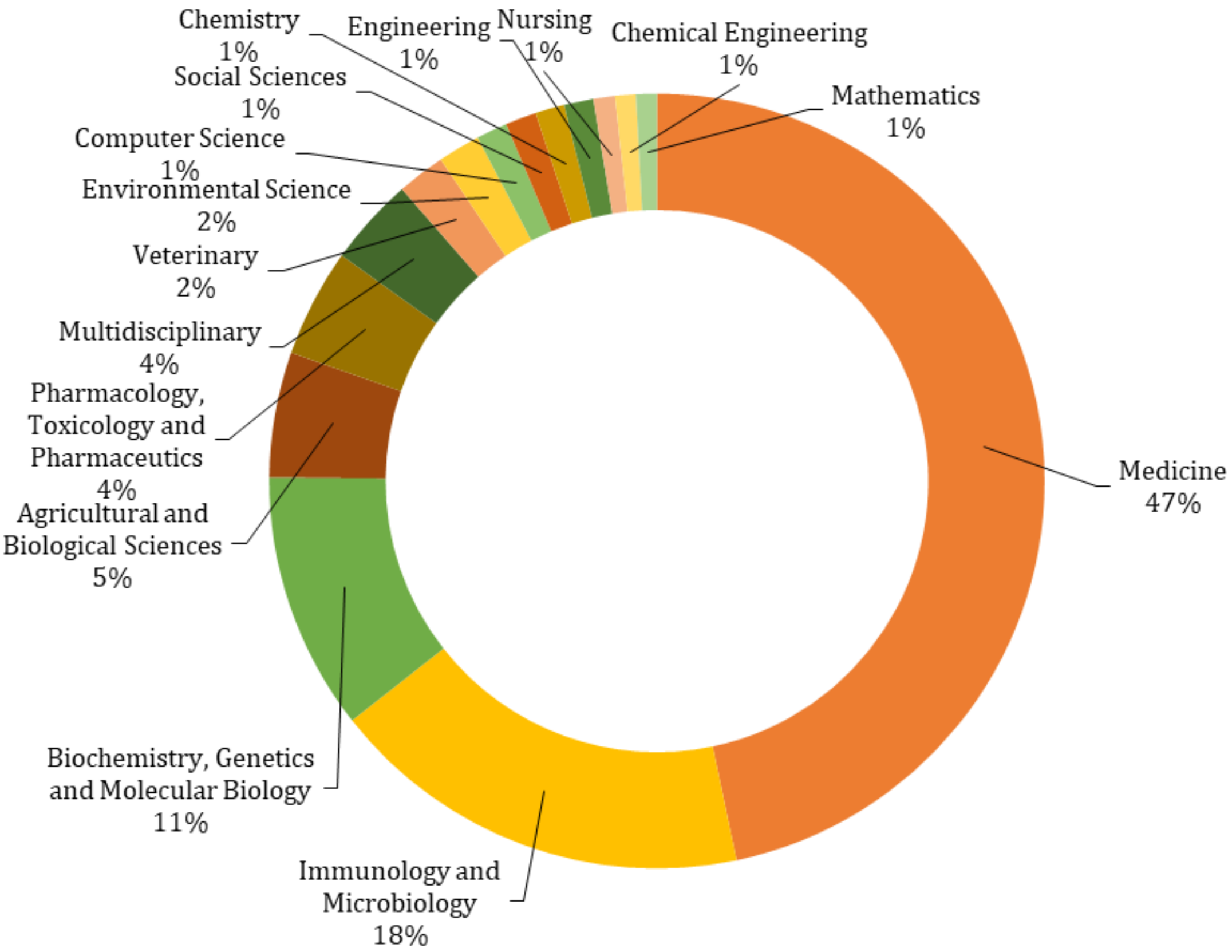

**Figure 2** Distribution of MERS studies across *Subject Areas*.

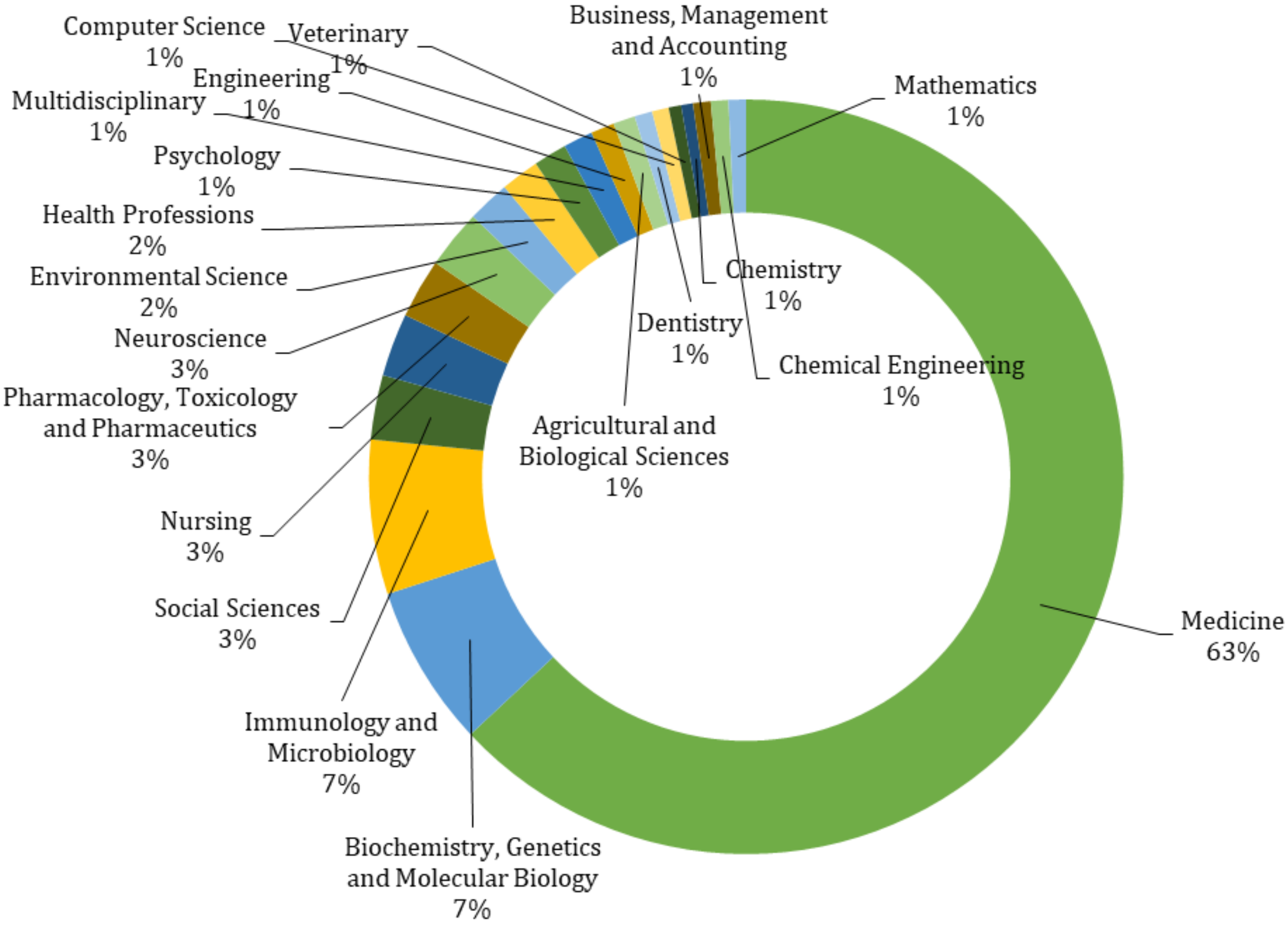

**Figure 3** Distribution of Covid-19 studies across *Subject Areas*.

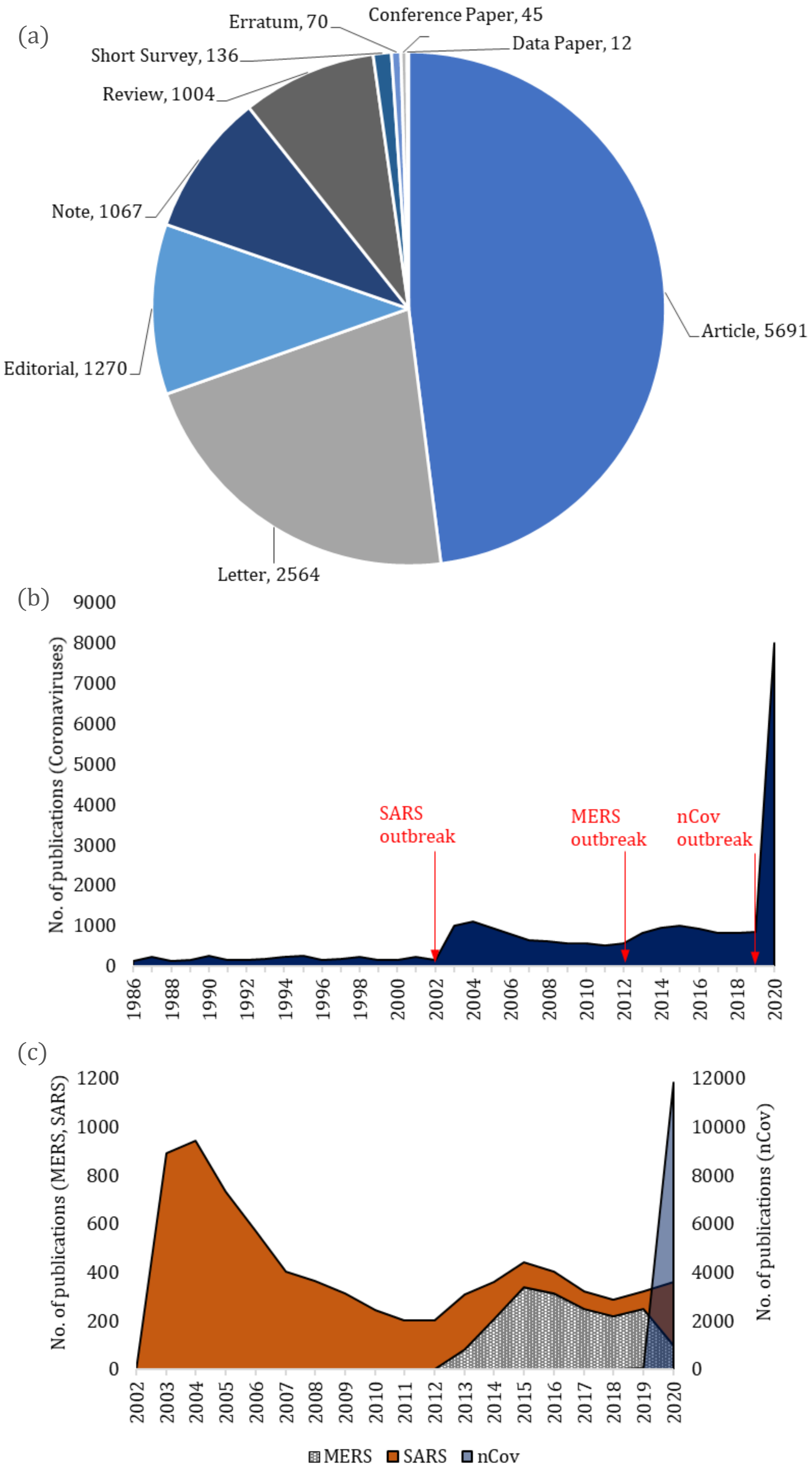

**Figure 4** (a) Distribution of Covid-19 studies across *Document Types*, (b) Temporal distribution of the number of Coronavirus studies, (c) Temporal distribution of the SARS, MERS and Covid-19 studies.

## 3. Keyword co-occurrence analyses

The co-occurrence of keywords associated with the SARS, MERS and Covid-19 literature were analysed using VOSViewer (Van Eck and Waltman 2010). Each analysis was performed on the separate set of data associated with the literature of interest. The maps of keyword co-occurrences associated with SARS, MERS and Covid-19 literatures are provided in Figures 5, 6 and 7 respectively. The minimum number of occurrences for the keywords to be included in the map was set to 5 in all three cases. The unit of analysis has also been set to *all keywords* (that includes both *author* and *index* keywords) and the method of counting was *full counting*. Figures A1 and A2 in the Appendix illustrate the map associated with the SARS literature overlaid respectively with the average year of publication and average number of citations associated with the studies where these keywords have occurred. Figure A3 and A4 present the counterpart outputs for the MERS literature analysis. Figure A5 is a heatmap of Covid-19 keyword co-occurrence and Figure A6 overlays the Covid-19 keywords map in Figure 7 with the colour-coding of the average number of citations. Given that almost all studies of Covid-19 are 2020 items, the colour-coding related to the average publication year was forgone in regard to this literature. Maps of term occurrences based on the analysis of the title and abstract of studies on SARS, MERS and Covid-19 have also been presented in Figures 7, 8 and 9 respectively. While the below analysis focuses mainly on the interpretation of the keyword maps, similar patterns are by-and-large observable through analysis of the title and abstract terms of these studies.

With respect to each of the three literatures, three distinct clusters of keywords were identifiable. These clusters showed certain patterns of commonality across the three datasets. Each map presents a distinct cluster of keywords that seem to be associable to the studies related to **public health emergency management and the prevention of epidemic**. This would be the red cluster in Figure 5 (SARS), the green cluster in Figure 6 (MERS) and the red cluster in Figure 7. Here, this is referred to as **Cluster (i)**. In this cluster, one can observe terms such as those associated with **general public health** including "*wold health organisation*", "*public health*", "*public health service*", "*global health*", as well as those associated with **disease outbreaks** including "*emergency*", "*health risk*" "*epidemics*", "*pandemic*", "*outbreak*", "*viral diseases*", "*virus infection*", "*communicable disease*", "*transmission*", "*travel*". Terms representing measures of **emergency severity** also appear in this cluster including "*mortality*", "*fatality*", "*morbidity*", "*infection risk*". This cluster also includes terms that are linked to the **prediction of disease propagation**. These are terms such as "*mathematical model*", "*modelling*", "*simulation*", "*statistical model*" and "*prediction*" that have commonly occurred in this cluster. The cluster includes terms affiliated with measures of **disease control and spread prevention** such as "(*social/patient) isolation*", "*quarantine*", "*hygiene*", "*handwashing*", "*prevention*", "*infection control*", "(*population) surveillance*", "*mass screening*", "*(face) mask*", "*contact tracing*". The cluster also represents keywords that attributable to **public policy making** and **social protection** such as "*health care planning*", "*health care policy*", "*health care quality*", "*leadership*", "*disaster planning*", "*polices*".

The **Cluster (i)** of keywords also have distinctly and commonly across all three datasets represented keywords that are attributable to the studies on **mental health** impact of the epidemic. These are keywords such as "*mental health (service)*", "*psychiatry*", "*psychology*", "*mental stress*", "*anxiety*", "*fear*", "*mental disease*". These studies have often used methods such as "*questionnaire(s)*" and "*survey(s)*" that have commonly reflected in this Cluster across the three literatures. Issues surrounding the safety of **medical facilities** and **medical staff** also appear to have been addressed mainly by studies whose keywords are attributable to this cluster. These studies have generated keywords such as "*health care personnel*", "*nurse(s)*", "*medical staff*", "*hospital*", "*health care facility*", "*personal protective equipment*" that are distinctly observable in **Cluster (i)** of keywords across all three datasets.

The **economic** aspects of the epidemics also seem to have been addressed particularly by Covid-19 as reflected in **Cluster (i)** of the Covid-19 literature**.** These have been reflected in terms such as "*economics*", "*economic aspect*", which have occurred frequently enough in Covid-19 studies for them to appear distinctly on the map. The trace of such cohort of studies is, however, not as clearly identifiable based on the SARS and MERS maps as is it with respect to the Covid-19 literature. This could be explained by the greater magnitude of the societal impact of Covid-19 outbreak compared to SARS and MERS.

The names of the **countries** and **regions** have almost invariably appeared in **Cluster (i)** across all three datasets. In certain cases, the country names that have occurred most are those from which the outbreaks originated or those that suffered most from the impact of the outbreak. For example, "*Saudi Arabia*" appears quite distinctly on the **Cluster (i)** of the MERS dataset. Similarly, the occurrence of the names of south-east Asian countries/regions such as "China", "Hong Kong", "Taiwan", "Singapore" on the **Cluster (i)** of the SARS map, or the term "Wuhan" on the **Cluster (i)** of the Covid-19 map are quite notable. The occurrence of the name of the countries also could be a reflection of the early studies with respect to each outbreak that have addressed the local impacts/spread of the outbreaks on their own society. On the issue of **early studies**, the terms "*letters*", "*editorial*", and "*review*" (which have intentionally been kept on the maps) seem to also have distinctly occurred in **Cluster (i)** of each literature which is another indication that this cluster includes early studies that appeared at a time where the amount of data and clinical trials were insufficient for full-length articles. An inspection of the Figures A1 and A3 does, in fact, confirm this hypothesis at least in association with the SARS and MERS literature, that the **Cluster (i)** of keywords represent studies that on average emerged early during the developments of their respective literatures. Figures A2, A4 and A6 that have illustrated the colour-coding of the average number of citations on the maps also show that, although **Cluster (i)** is associated with the early studies that generally predated studies of the two other clusters and although it represents the largest variety of topics compared to the two other clusters, it is also associated with studies that, on average, been the recipient of a lesser number of citations when compared to the two other clusters. This pattern also appears to have commonly occurred across all three datasets.

A second cluster of keywords associated with each of the three literatures were also discovered that is attributable to the studies on the **chemistry and physiology of the virus**, or **viral pathogenesis,** or in other words, the **chemical constitution** of the virus (Knight 1954), a part of **virology** that investigates the biological processes and activities of viruses that take place in infected host cells and result in the replication of a virus. For the SARS map in Figure 5, as well as the Covid-19 map in Figure 7, this would be the green cluster, whereas for the MERS map (Figure 6), this cluster is red. According to the maps, the most distinct terms associated with this cohort of virology studies on SARS, MERS and Covid-19 are terms such as "*virus protein*", "*virus entry*", "*chemistry*", "*metabolism*", "*physiology*", "*pathology*", "*cell line*", "*(virus/viral) protein(s)*", "*molecular model(s)*", "*virus genome*", "*virus rna*", "*virus replication*", "*mutation*", and "*enzyme activity*". As this sector of studies often use "*animal model(s)*", terms such as "*animal cell*", "*animal experiment*", "*controlled study*", "*mice*" and "*mouse*" have frequently appeared in **Cluster (ii)** associated with each of the three literatures. In reflection of the fact that these cohort of studies ultimately seek "*drug design*", in addition to generic common terms such as "*drug design/potency/structure/synthesis*", the names of the specific potential drugs that have been investigated in relation to each disease have appeared in this cluster. This includes terms such as "*hydroxychloroquine*" or "*remdesivir*" on the Covid-19 map. An inspection of the maps overlaid with the average year of publications for SARS and MERS in Figures A1 and A3 in the Appendix suggests that, on average, this cohort of studies are generally the last to emerge in the published domain compared to the two other major clusters, but they receive relatively high citations on average (according to Figures A2, A4 and A6).

A third and relatively smaller cluster of keywords was commonly identifiable in relation to each three literatures. This cluster has been visualised in blue colour across all three maps of keyword co-occurrence. The studies represented by this cluster of keywords, here referred to as **Cluster (iii)**, appear to have been more closely focused on the developments of **antibodies** and **vaccines**. The terms "*treatment*", "*treatment outcome*", "*disease severity*", "*antiviral therapy*", "*prognosis*", "*drug safety*", "*prospective/retrospective study*", "*immunology*", "*immunotherapy*", "*innate immunity*", "*immune response*", "*virus/viral vaccine(s)*", "*virus/viral antibody*" are notable across these studies. Terms affiliated with studies related to **treatments** and **clinical care** of respiratory disease patients also appear in this cluster. This includes terms such as "*artificial ventilation*", "*intensive care unit*", as well as **symptom** and **organ** terminologies associated with each disease, terms such as "*fever*", "*headache*", "*diarrhea*", "*lung (injury)*", "*coughing*", "*liver injury*", "*kidney*". Terms affiliated with **cohort analysis** studies have appeared in this cluster of the maps associated with each literature. This is reflected in terms such as "*female*", "*male*", "*child*", "*infant*", "*young adult*", "*adult*", "*age*", "*middle aged*", "*pregnant*", "*pregnancy*". This pattern of the cohort analysis keywords appearing in Cluster (iii) is particularly common across the MERS and Covid-19 studies. For SARS, these terms have largely appeared in the red cluster, at the border between the red and blue clusters.

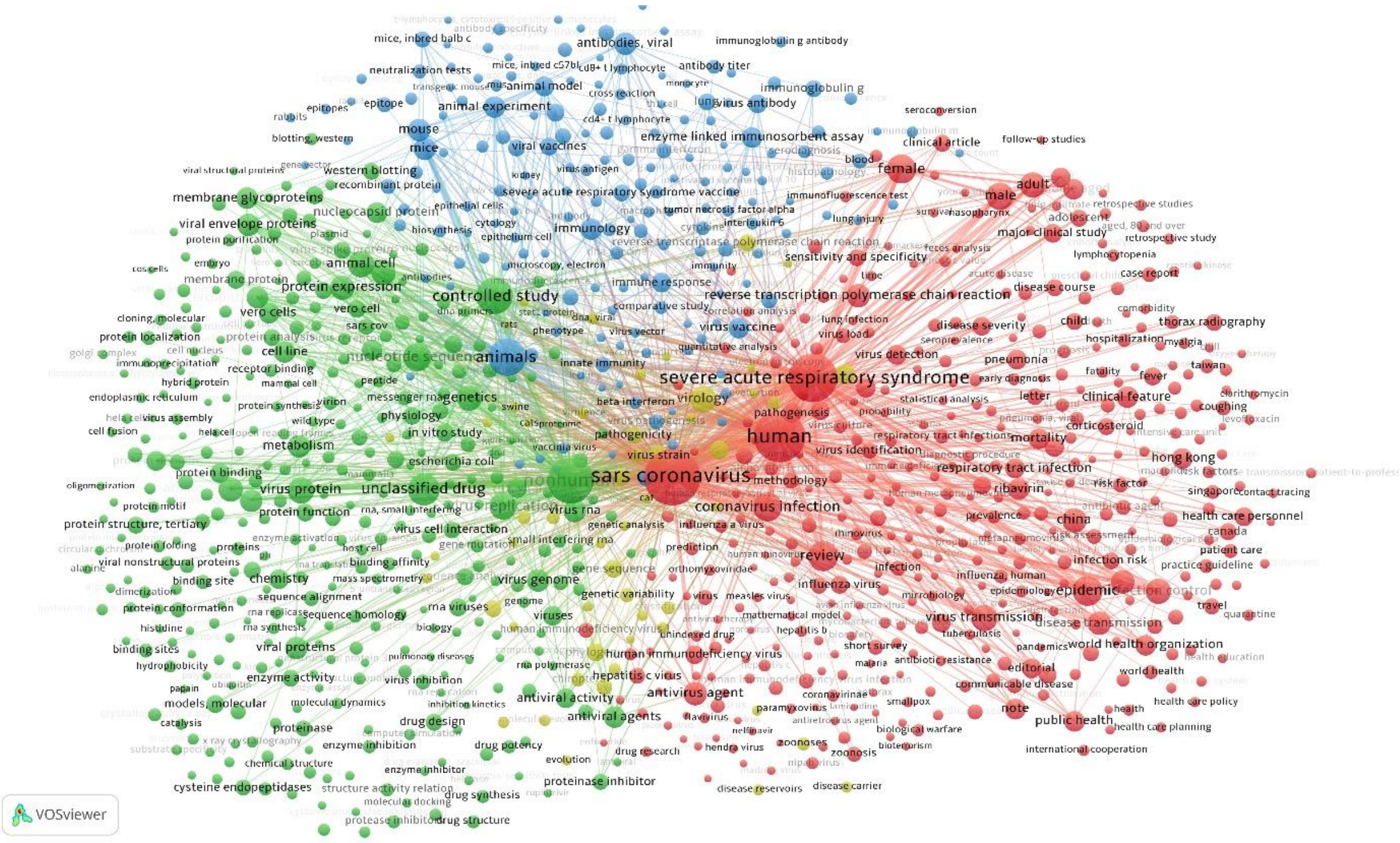

**Figure 5** The map of keyword co-occurrences associated with the SARS literature.

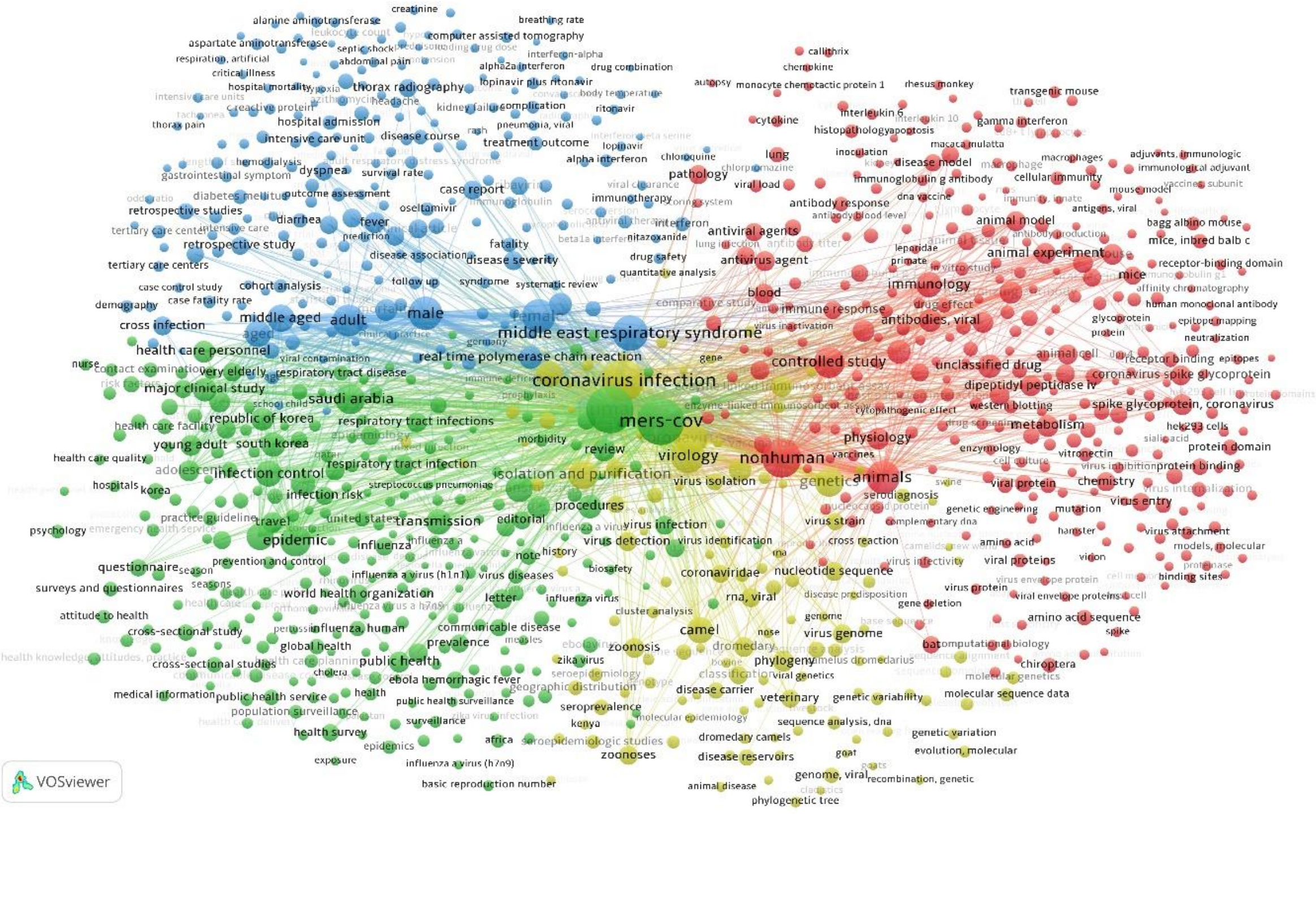

**Figure 6** The map of keyword co-occurrences associated with the MERS literature.

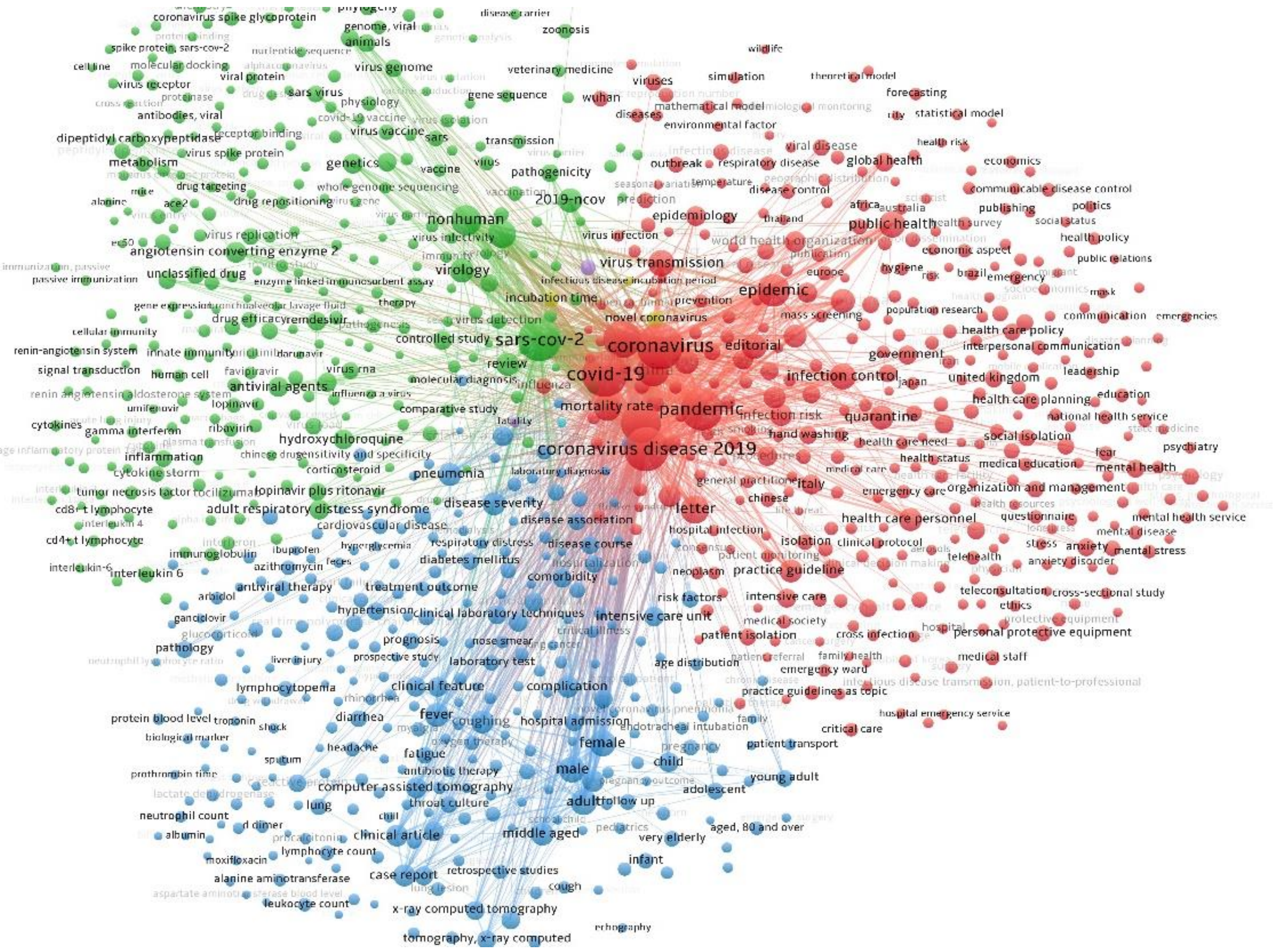

**Figure 7** The map of keyword co-occurrences associated with the Covid-19 literature.

## 4. Bibliographic coupling and citations of journals

Bibliographic coupling of the studies on SARS, MERS and Covid-19 were analysed at the level of their sources/journals. Figure 8, 9 and 10 show the maps of journal bibliographic coupling associated with SARS, MERS and Covid-10 literatures respectively. The node sizes are proportional to the number of documents published by the corresponding sources and the thickness of the links are proportional to the degree of bibliographic couplings between the sources connected by each link. The minimum number of documents associated with each node/journal to appear on the map has been set to 10. No minimum strength was set for links to be visualised on the map.

A first-glance comparison shows that while the maps associated with SARS and MERS are well connected, connections across the Covid-19 map are rather sparse. Both the SARS and MERS maps include three major and distinct clusters of bibliographically coupled journals in addition to one minor and smaller cluster. These clusters show relatively strong degrees of inter-connectivity, whereas, this feature is not shared by the Covid-19 map. The observation is understandable in light of the fact that the SARS and MERS literatures are relatively well established and have each been under development over a period of several years, whereas the Covid-19 literature is an emerging field, and newly published studies do not seem to be sharing many references as of yet. The comparison also suggests that the Covid-19 studies are generally scattered across a broader variety of journals and subject areas, as opposed to the SARS and MERS publications that seem to have been concentrated across a smaller set of specialty journals. This is also consistent with our observations from Figures 1-3 showing that studies of Covid-19 are scattered across a broader variety of subject areas compared to the SARS and MERS literature. Though not shown in Figure 3, due to the respective values being smaller than 1%, journals in the following subject areas (that are deemed minor areas in relation to Covid-19 literature) have each published a relatively considerable number of studies on this topics (a phenomenon that is not necessarily common with respect to the literature of other Coronaviruses): **Arts and Humanities** (110 items[1], where the most active journal has been *Social Anthropology* (24 items) covering topics such as "climate change reactions" (Bychkova 2020), or "legal voids linked to declared states of emergency" (Karaseva 2020)), **Economics, Econometrics and Finance** (84 items, with *Economic and Political Weekly* (36 items) being the most active journal of that category, covering topics such as "food supply chains" (Reardon *et al.* 2020), "economic stimulus packages" (Mulchandani 2020) or "reverse migration" (Dandekar and Ghai 2020)), **Physics and Astronomy** (77 items, where *Chaos Solitons and Fractals* (16 items) has been the most active publication outlet, covering topics such as "mathematical models for forecasting the outbreak" (Barmparis and Tsironis 2020, Bekiros and Kouloumpou 2020, Boccaletti *et al.* 2020, Ndaïrou *et al.* 2020, Postnikov 2020, Ribeiro *et al.* 2020, Zhang *et al.* 2020)), **Energy** (67 items, with *International Journal of Advanced Science And Technology* (44 items) being the most active journal in that category, covering topics such as "Flexible work arrangement in manufacturing" (Sedaju *et al.* 2020)), **Material Sciences** (57 items, with *ACS Nano* (10 items) being the most active outlet in that category, covering topics such as “3-D printed protective equipment” (Wesemann *et al.* 2020)), **Decision Sciences** (23 items, with *Lancet Digital Health* (8 items) and *Transportation Research Interdisciplinary Perspectives* (4 items) being the most active outlets in that category, covering topics such as “the effect of social distancing on travel behaviour” (De Vos 2020) or “the implementation of drive-through and walk-through diagnostic testing” (Lee and Lee 2020)), **Earth and Planetary Sciences** (22 items, with *Indonesian Journal of Science and Technology* (8 items) being most active in that domain, covering topics such as “the deployment of drones in sending drugs and patient blood samples” (Anggraeni *et al.* 2020)).

[1] Note that these Figures are based on a renewed Scopus search on May 30, 2020 when the total number of Covid-19 studies had already exceeded 13,700.

An evident source that seem to have consistently published a substantial portion of studies on SARS and MERS is *Journal of Virology*. This journal, however, has not published a considerable number of studies on Covid-19, and with only eight publications on this topic at the time of writing this article, it does not have a strong representation on the map associated with Covid-19 literature. *The Lancet* and *Science*, however, are two major outlets notably observable on all three maps. For Covid-19 studies in particular, journals such as *Journal of Medical Virology*, *The BMJ*, *The Lancet*, *Journal of Infection*, *Science*, *Nature*, *Science of the Total Environment* and *Medical Hypotheses* have been most notable outlets of publications so far. Some of these outlasts, such as *Science of the Total Environment* and *Medical Hypotheses*, do not have a strong representation on the maps associated with the SARS or MERS.

In terms of the bibliographic coupling of the sources for SARS publications, the strong relation between *Journal of Virology* and *Virology*, and to lesser extent, with *Emerging Infectious Diseases* are outstanding. For MERS publications, the strong bibliographic coupling of publications between *Emerging Infectious Diseases* and *Journal of Virology* is most outstanding. For Covid-19 publications, the one outstanding bibliographic coupling relation is between *Journal of Medical Virology* and *Journal of Infection*.

The analyses of journal citations also showed similar patterns of scatter and relatively unclear clusters in relation to the Covid-19 literature compared to well-defined clusters of journal citations for SARS and MERS literatures. Consistent with the previous observation with respect to journal bibliographic coupling, the Covid-19 literature seems to be also much less cohesive in terms of its journal citation networks, when compared to the SARS and MERS literatures. As discussed earlier in relation to bibliographic couplings, this could also be partly explained by the fact that Covid-19 papers are scattered across a more diverse range of journals and broader variety of subject categories. For the maps of journal citation relations presented in Figures 11, 12 and 13, only the strongest citation relations have been visualised. They have also been overlaid with average citation colour coding. For SARS, the maps present very strong citation relations between publications in *Journal of Virology* and those of *Virology*, *PNAS* and *Science* (and to a lesser extent, with *Emerging Infectious Diseases*, *The Lancet* and *New England Journal of Medicine*). For MERS publications, the strong citation relation between *Emerging infectious Diseases* and *New England Journal* of Medicine are the ones that most stand out. Also, for the Covid-19 literature, three pairs of strong citation relations are identifiable: first one between *Journal of Medical Virology* and *The Lancet*, second one between *Journal of Medical Virology* and *New England Journal of Medicine*, and third one between *European Radiology* and *Radiology*.

In Figures A10-A14 in the Appendix, the nodes of the bibliographic coupling maps have been colour-coded by the average year of publications and the average citations per document associated with the journals that each node represent (except for the Covid-19 map that has only been overlaid with the average citations). According to these maps, *Emerging Infectious Diseases* and *The Lancet* have been a major source of publications for early studies on both SARS and MERS. This pattern for *The Lancet* seems to have extended to Covid-19 studies as well, as this journal has published a substantial portion of early studies on this topic. For SARS, the strong representation of *Chinese Medical Journal* and *Chinese Journal of Microbiology and Immunology* among the journals that published early studies are notable, a pattern that could be explained by the geographical origin of the SARS outbreak. Such pattern is to a less obvious extent observable in regard to the MERS literature through representations of outlets such as *Eastern Mediterranean Health Journal* and *Saudi Medical Journal* on the bibliographic coupling map associated with this literature by colours associated with relatively early publications.

In terms of the average number of citations, studies published by *The Lancet* have consistently and across all three literatures received high citations. For the SARS literature in particular, other sources whose publications on this topic received high numbers of citations are *New England Journal of Medicine*, *PNAS*, *Current Medical Chemistry*, *Journal of Pathology*, *Plos Biology*, and *Nature Reviews Microbiology*. The counterpart outlets associated with the MERS literature are *Journal of Virology*, *Science*, *mbio*, and *New England Journal of Medicine*. For Covid-19 studies, publications of *Nature*, *Eurosurveillance*, *The Lancet Infectious Diseases*, *JAMA*, *The New England Journal of Medicine*, and *Radiology* have on average been among the most cited studies.

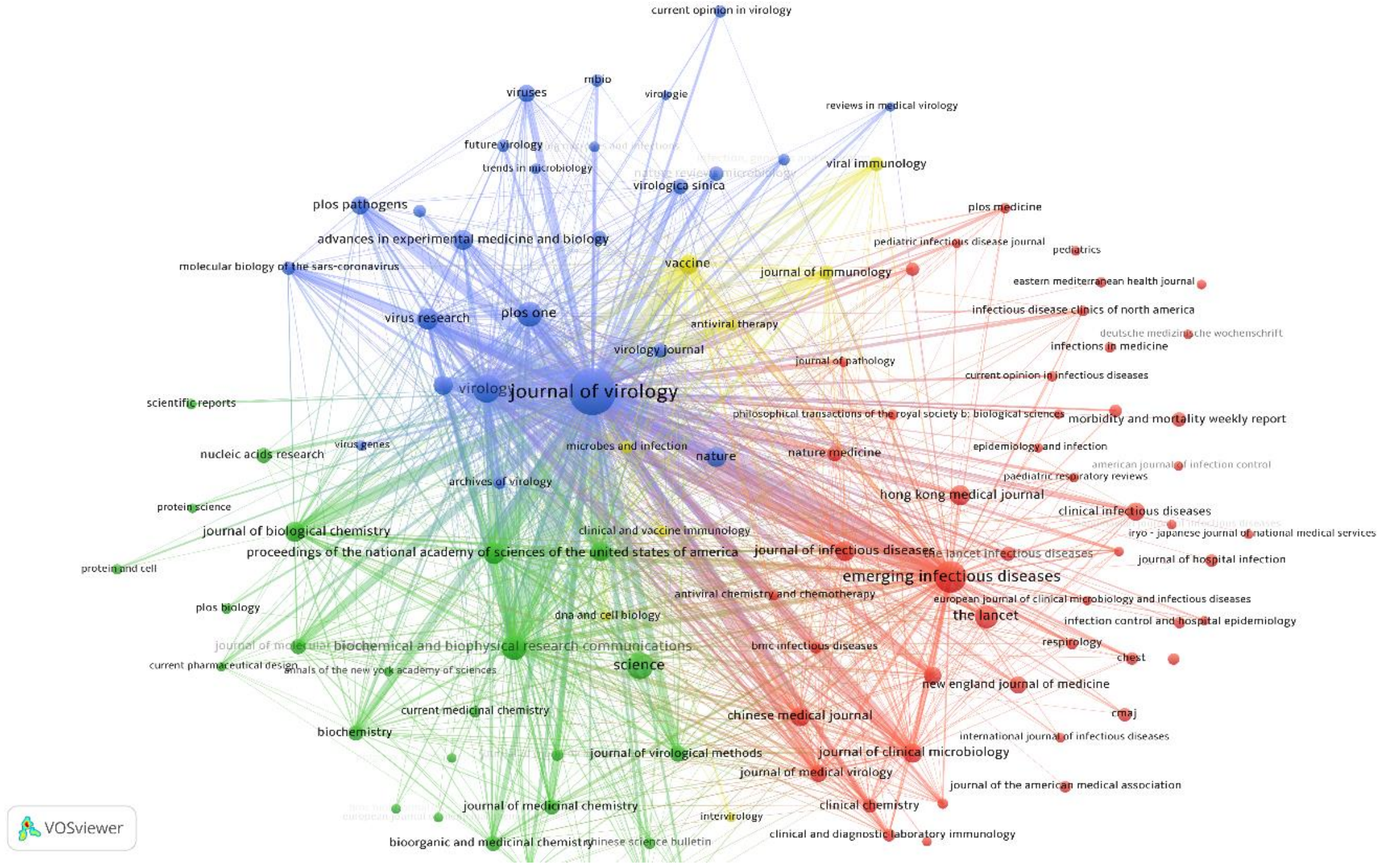

**Figure 8** The map of bibliographic coupling of sources associated with the SARS literature.

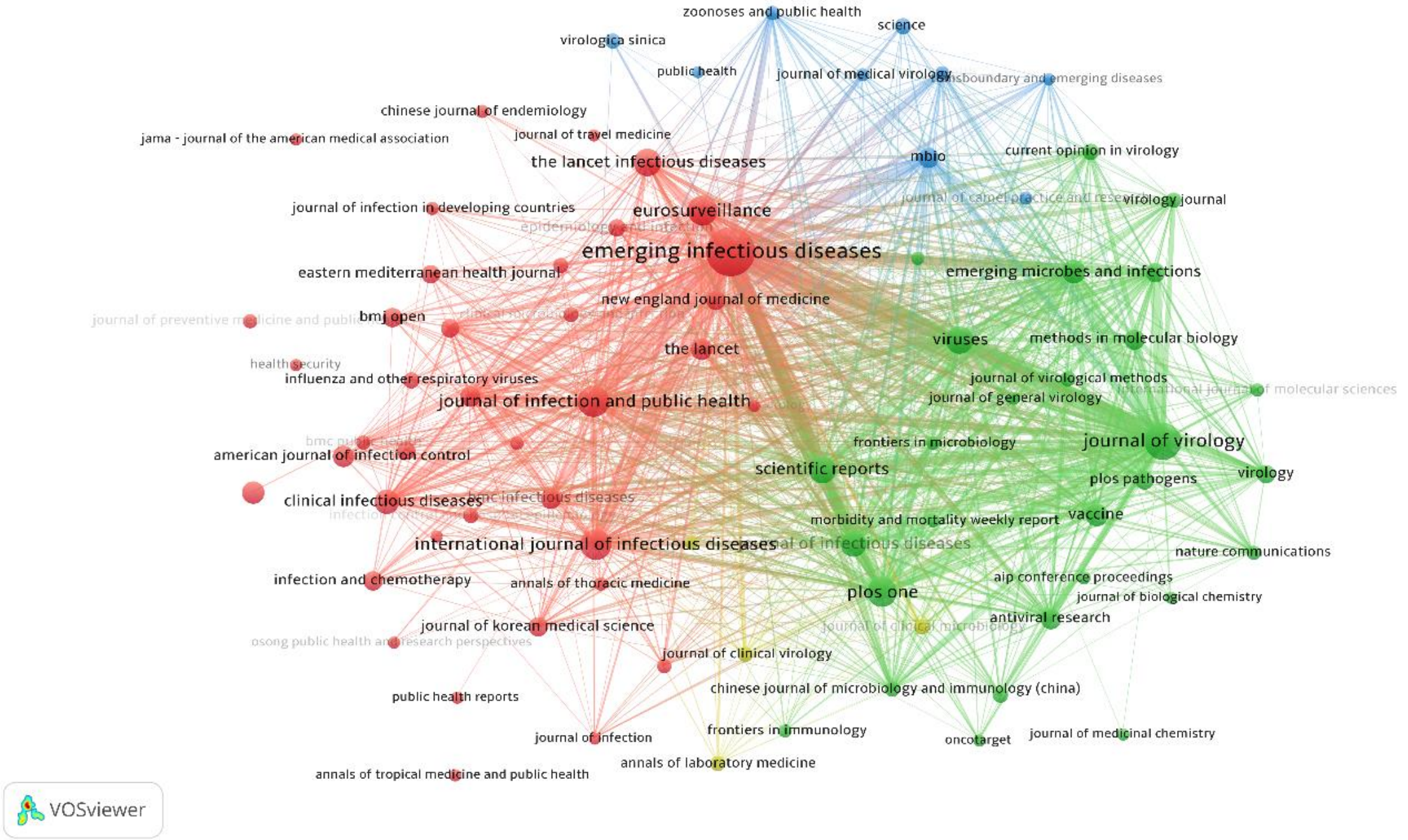

**Figure 9** The map of bibliographic coupling of sources associated with the MERS literature.

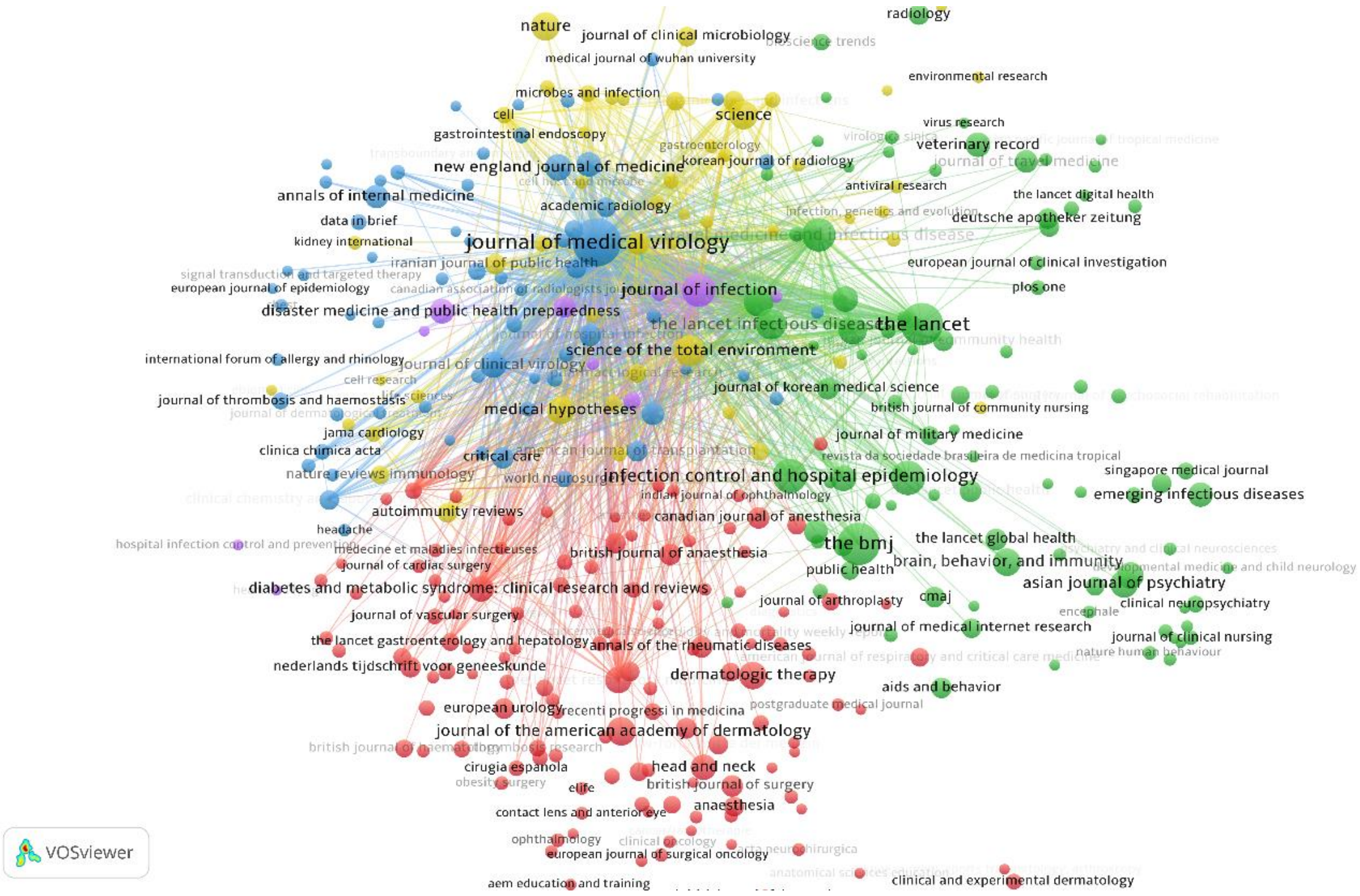

**Figure 10** The map of bibliographic coupling of sources associated with the Covid-19 literature.

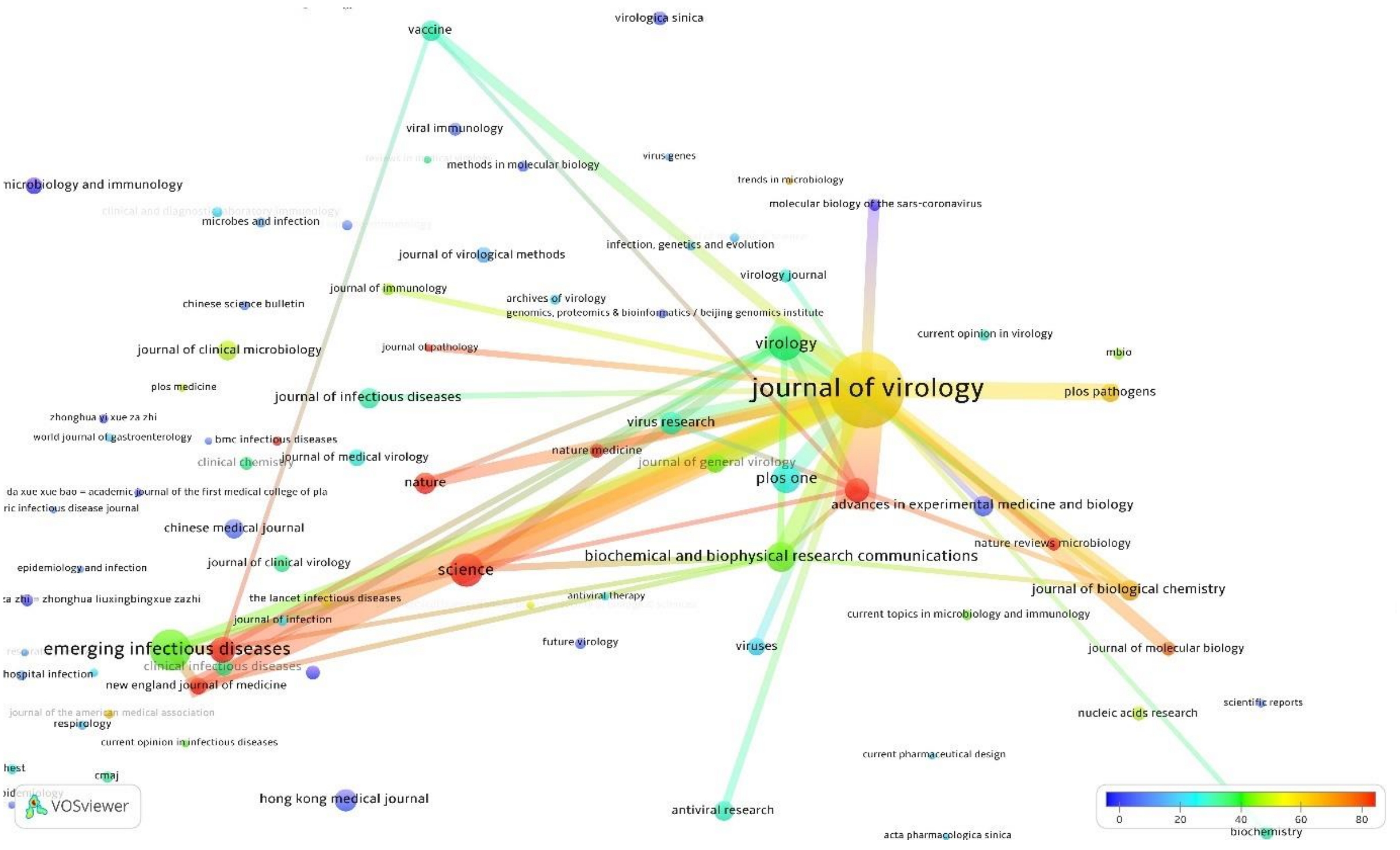

**Figure 11** The map of strongest citation network of sources for SARS overlaid with the colour-coding of the average citation per document

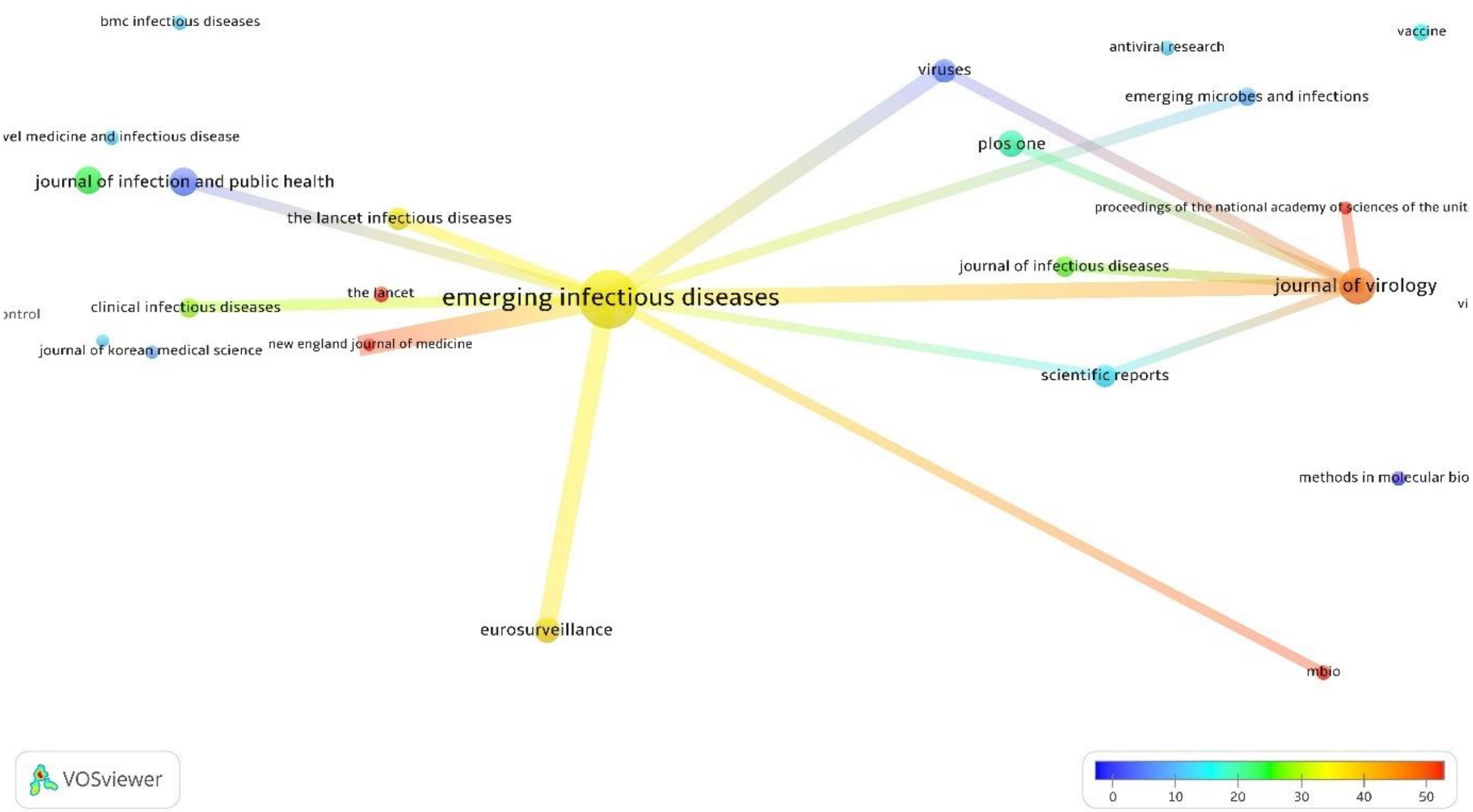

**Figure 12** The map of strongest citation network of sources for MERS overlaid with the colour-coding of the average citation per document

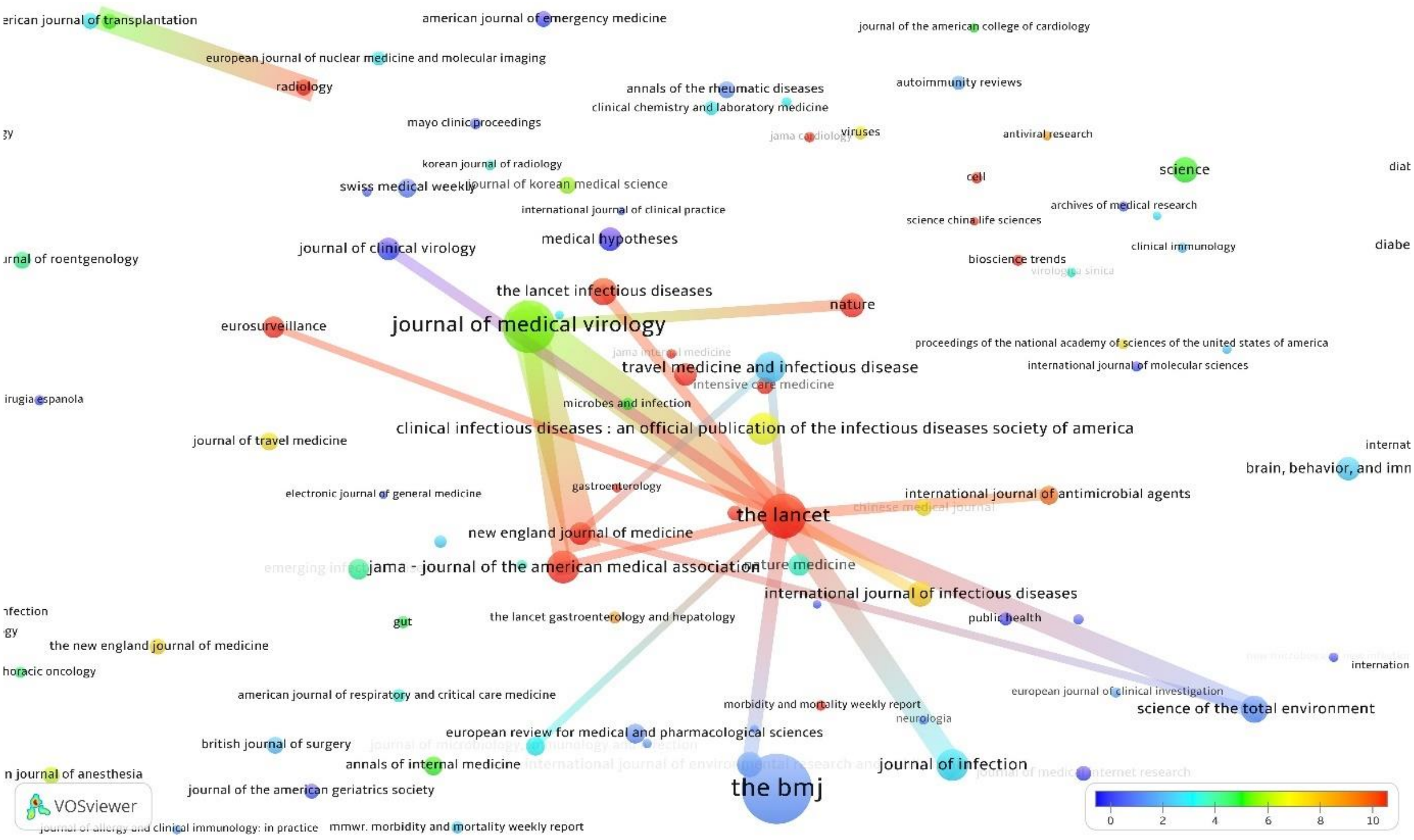

**Figure 13** The map of strongest citation network of sources for Covid-19 overlaid with the colour-coding of the average citation per document

## 5. Co-authorship between countries

Collaborations of authors aggregated at the level of the countries were also analysed with respect to the SARS, MERS and Covid-19 literatures. Outputs of the analysis are presented in Figures 14, 15 and 16 for SARS, MERS and Covid-19 respectively. In each map, the size of nodes, each corresponding with a country, are proportional to the number of published documents with an author affiliated with the institutes of those countries. The links connecting the nodes indicate co-authorships between authors residing in the countries, while the thickness of the links represent the strength (i.e. frequency) of the co-authorships. The colour assigned to each node represents the average number of citations that documents authored by the countries have received. The minimum number of documents for country names to appear on the maps has been set to 5.

Comparison across the three maps of co-authorships shows a pattern of author involvement from the regions where each viral outbreak originated. Studies authored by researchers affiliated with Chinese institutes are well represented in all three cases, but clearly more so with respect to the SARS and Covid-19 literature, diseases whose first cases were recorded in China. The involvement of Chinese authors is relatively less notable in relation to the MERS studies whose origin was in the Middle East. Instead, with respect to the MERS literature, it appears that authors affiliated with institutes in Saudi Arabia have been exceptionally overrepresented in the publications. This is also, to a lesser extent, the case with Egypt being notably presented on the MERS map of the country co-authorships.

On SARS studies, Chinese authors have most strongly collaborated with authors residing in the United States, followed by authors from Germany, Taiwan, Singapore, Japan, France, Australia, United Kingdom and Canada. The SARS network of collaborations for the authors affiliated with Institutes in the United States has been, by and large, similar to that of China, except South Korea, The Netherlands, Italy and Spain are also strongly represented in the collaborations with the United States.

The map of co-authorships associated with the MERS literature presents the names of many Middle Eastern countries including Saudi Arabia, Egypt, Lebanon, Iran, Tunisia, Qatar, Oman, Jordan, and United Arab Emirates which is a clear indication into the exceptionally strong representation of the authors from this region in these studies. The strongest network of co-authorships on this topic are observed between authors from the United States and those of Saudi Arabia, followed by China, United Kingdom, Egypt, South Korea, Canada and the Netherlands. The closest collaborators of Chinese authors on this topic, after the United states, have been from Saudi Arabia, United Kingdom and Egypt. The closest collaborators of Saudi Arabia, after the United States, have been Egypt, China and United Kingdom.

The Covid-19 map presents a considerably higher number of country names compared to that of the SARS and MERS literatures. It clearly shows that authors from more countries have become involved in studies of Covid-19, compared to the research on SARS and MERS that have apparently engaged a lesser number of countries. China, on the topic of Covid-19, has a very well spread and rather more evenly distributed network of collaborations with countries across the world, when compared with its network of collaboration on SARS and MERS. While its strongest collaboration has been with the United States, the names of many other countries appear on its network with no particular country standing out distinctly. Italy, as a country that was highly affected by the viral outbreak, has been exceptionally well represented on this map with a very strong link of collaboration with the United States, followed by United Kingdom at a smaller scale. This pattern of unique over-representation has to a lesser extent extended to Iran, Spain, France and Brazil as other countries also severely affected by the Covid-19 outbreak at early stages of the global spread.

The SARS studies with involvements of the authors affiliated with the institutes in the Netherlands have on average received the highest number of citations and this is followed by authors from Germany, as two countries whose authors have both published considerable number of documents and received high number of citations at the same time. This pattern was, to some extent, repeated in relation to the MERS literature, with studies from the Netherlands, Germany and United Kingdom having received on average highest number of citations. For studies published on Covid-19, studies from China have so far stood out in terms of both the magnitude of research activities and the average number of citations.

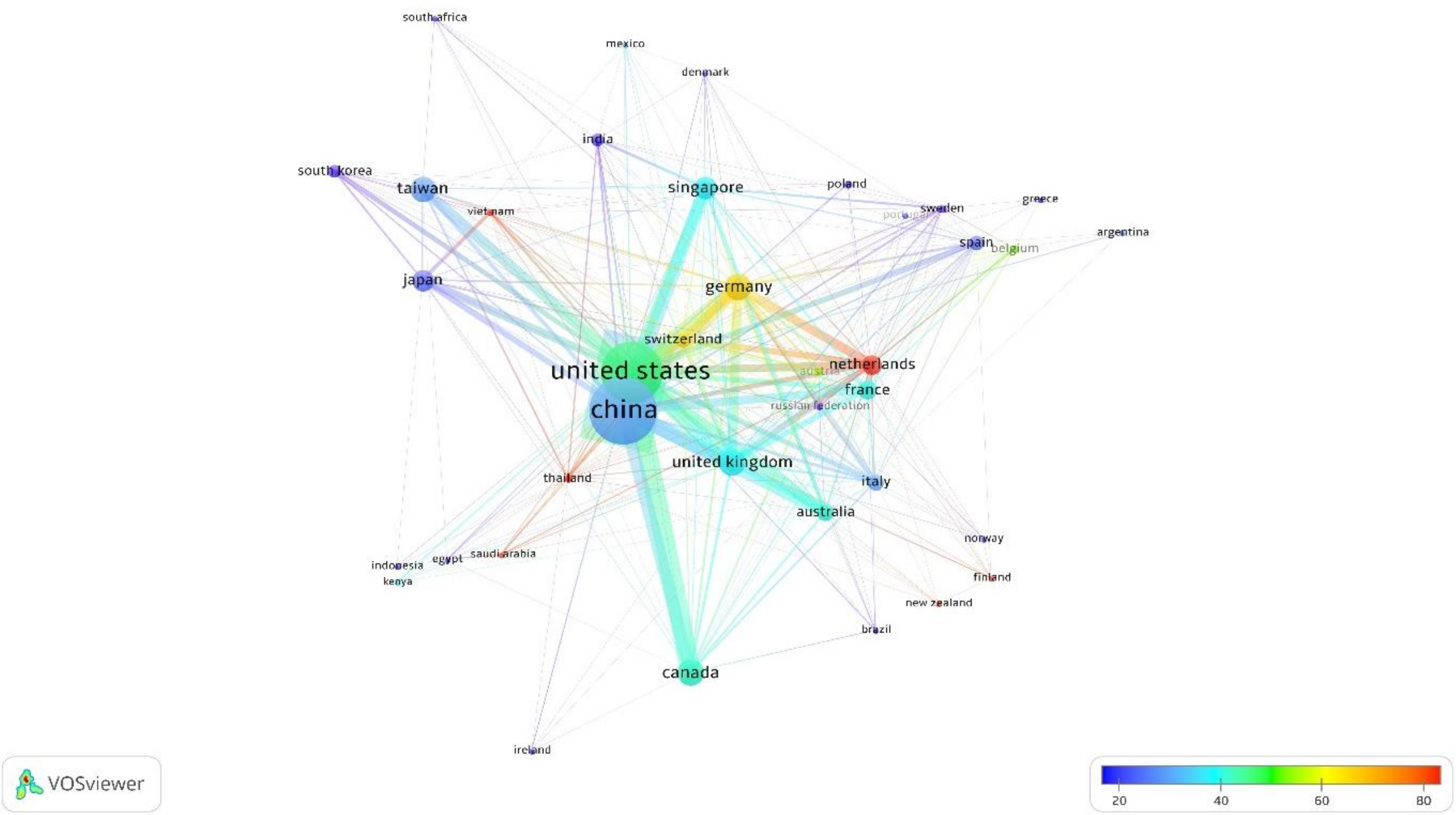

**Figure 14** The map of country co-authorships associated with the SARS literature.

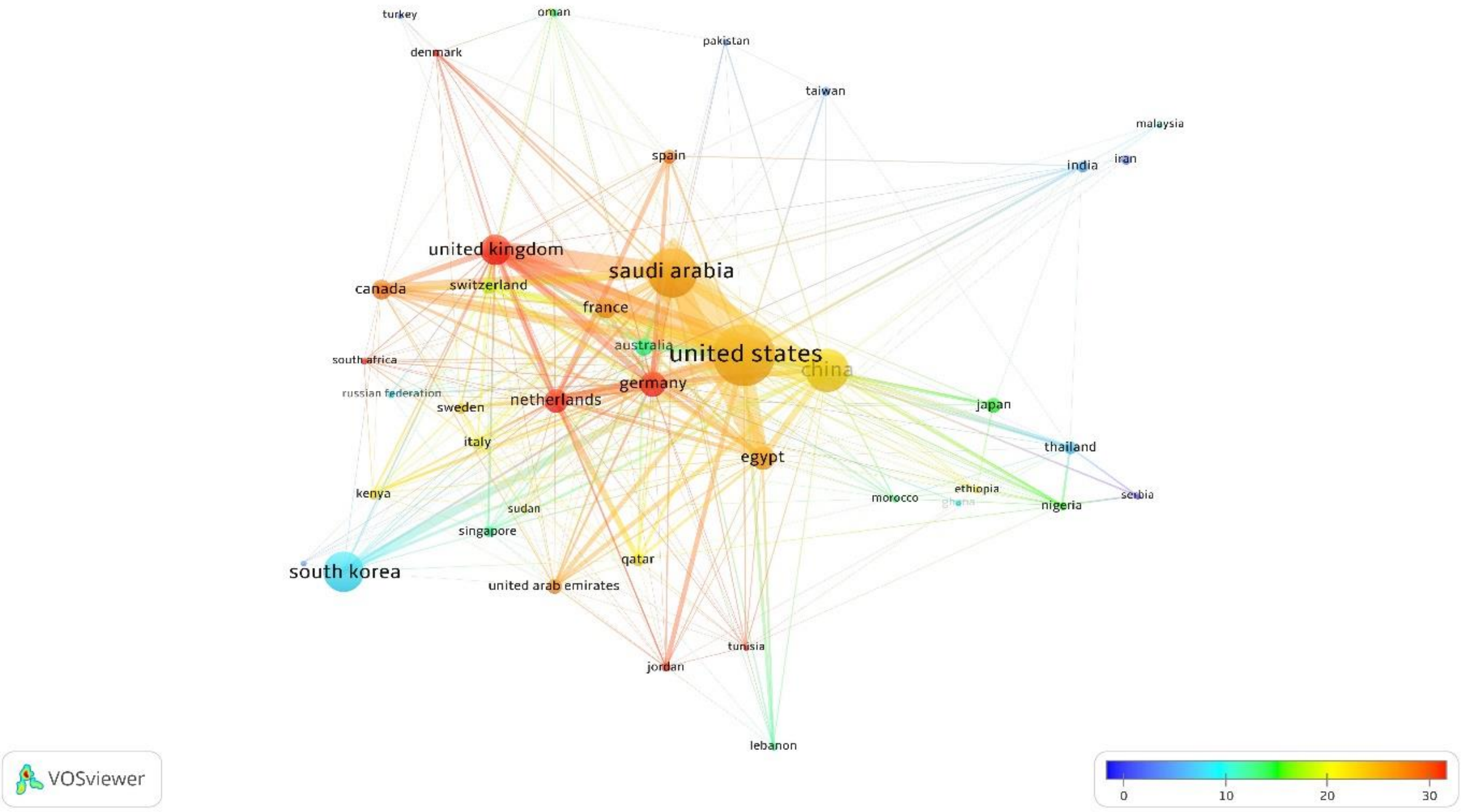

**Figure 15** The map of country co-authorships associated with the MERS literature.

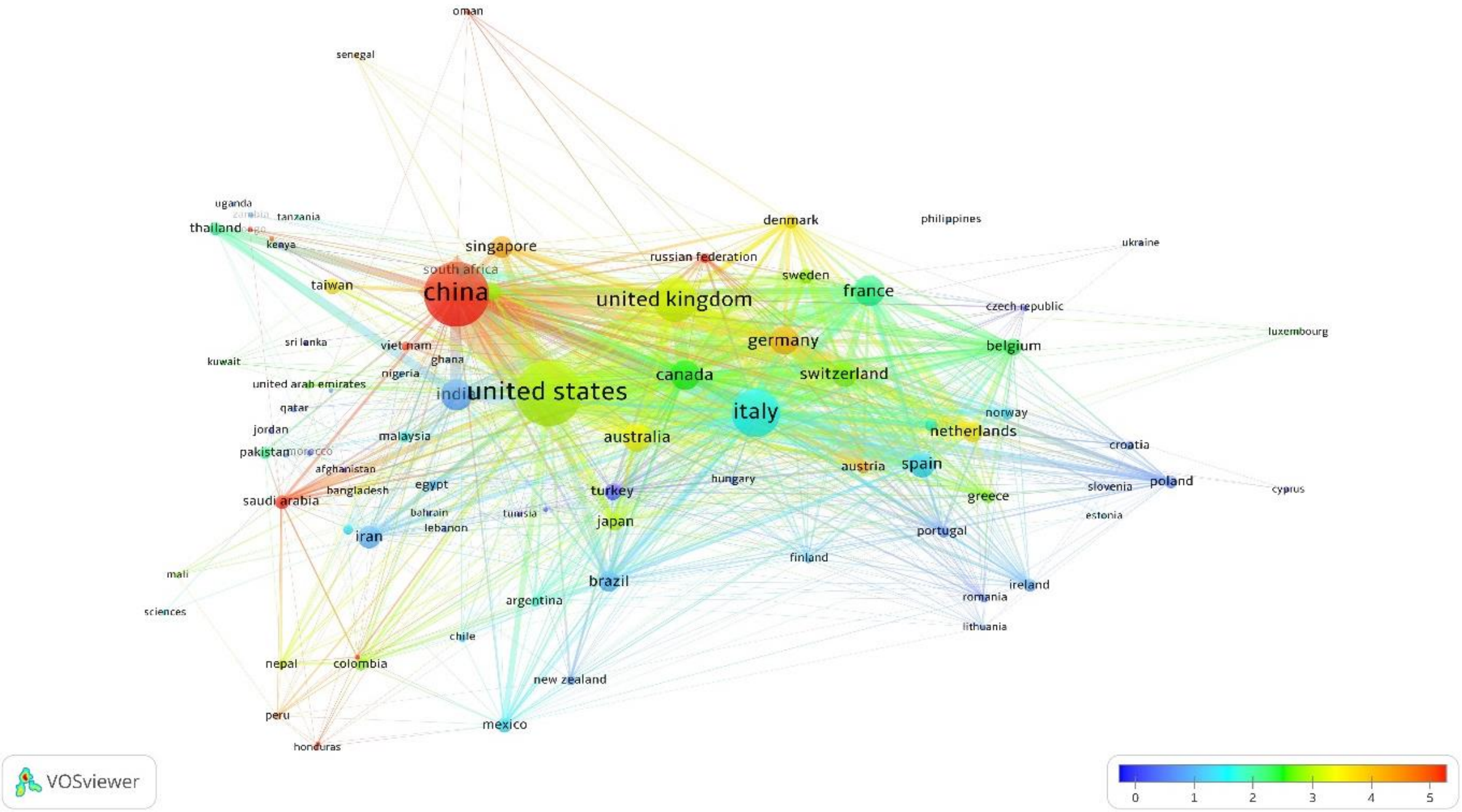

**Figure 16** The map of country co-authorships associated with the Covid-19 literature.

## 6. Summary and conclusions

Outbreaks of infectious diseases have often shown a pattern of generating a quick surge of publications on their respective topics, such that they often create an entirely new literature over a short amount of time (Olijnyk 2015, Tian and Zheng 2015). By all measures, however, the influx of research publications that began to emerge following the 2019 Novel Coronavirus outbreak outsizes those of the previous cases in the history of Coronaviruses, and perhaps arguably, in the history of infectious diaereses (Tian and Zheng 2015). This has certainly marked a new milestone in the timeline of research on Coronaviruses which dates back to 1968 (Almeida *et al.* 1968). According to the editor-in-chief of the *Journal of Virology*, as quoted in an article of *The Scientist* magazine (Jarvis 2020), this surge of research outputs has been to the extent that has inundated established Coronavirus researchers and domain experts with peer review requests to an extent that they are unable to cope. Parallel to such intensified efforts in the research, peer review and editorial fronts, widespread efforts are underway in synthesising, summarising and visualising these rapid developments, a pattern that has also been observed—though at much smaller scales—in relation to the previous epidemics of viral diseases (Kostoff and Morse 2011). In line with these endeavours, this work also aimed at quantifying and analysing scientometric aspects of the Covid-19 literature in contrast with those of the previous major Coronavirus diseases, i.e. SARS and MERS. The focus for sourcing these literatures has been on peer reviewed and published studies that have been indexed by Scopus.

An analysis of the timeline of the development of publications on Coronaviruses made clear that the SARS outbreak constitutes the first major milestone in the history of this research, an event that brought a then-unprecedented amount of attention to this topic. While Scopus has indexed a total of nearly 4,400 studies on Coronaviruses from 1968 till 2002, the three immediate years post the SARS outbreak (i.e. November 2002) have each recorded nearly 1,000 Coronavirus publications. This means that following the SARS outbreak, the then-36-year-old literature of Coronaviruses almost expanded by 70% within only three years. This trend, however, did not persist to the years succeeding 2005, as from that point on, a gradual decline in the rate of publications on Coronaviruses began. This continued until the 2012 MERS outbreak, another event that reinvigorated the coronavirus research, though not as substantially as that of the SARS epidemic. The pattern of a few years of increase in research activities on Coronaviruses followed by a gradual decline, observed in relation to SARS, also repeated in a similar fashion after the MERS outbreak. The outbreak of 2019 Novel Coronavirus, however, marks a unique milestone in this timeline. The magnitude of scholarly outputs prompted by this novel virus was to the extent that, in less than five months, 12,000 publications is already indexed by Scopus, a number that is nearly equivalent of 70% of the total amount of literature generated on Coronaviruses during the 50 years of this research prior to 2020.

By retrieving and disentangling the literatures linked to these three Coronavirus respiratory diseases, we sought to discover similarities and discrepancies of their research landscape from scientometric analysis perspectives. The most interesting pattern was the recurrence of three distinct clusters of studies in each literature as suggested by keyword co-occurrence analyses. It appeared that, following each outbreak, an early cluster of studies first emerged, addressing issues attributable to the public emergency management aspects of a pandemic, such as prediction of disease propagation, measures of outbreak control, public policy making, and concerns related to the protection of medical professionals and mental health. Compared to the two other clusters, studies of this cluster seem to have been the recipient of relatively smaller number of citations consistently across all three literatures. The two other clusters, one on virus chemistry and physiology and the other on vaccine, treatment and clinical care emerged relatively later but received higher number of citations.

Citation and bibliographic coupling analysis at the level of journals demonstrated that firstly, Covid-19 studies are scattered across a broader variety of sources and subject categories, and secondly, its network of journal relations is still not as cohesive as those of the SARS and MERS literatures. As the literature on Covid-19 further develops, more cohesive patterns of bibliographic coupling or journal citations might emerge. While *Journal of Virology* and *Emerging Infectious Diseases* seem to have been two major outlets commonly prominent across both SARS and MERS literatures, their presence in the literature of Covid-19 seem not be distinctly notable. Instead, a great portion of Covid-19 studies have concentrated across three journals: *The Lancet*, *The BMJ*, and *Journal of Medical Virology*. While major multidisciplinary journals, particularly *Science*, *Nature*, *PNAS* and *PLOS ONE*, have, to varying degrees, been active in publishing studies on all three topics, their influence is most notable in relation to the SARS literature where they have published a substantial portion of studies and those studies have been recipients of relatively high number of citations too.

The involvement of authors from various countries on the publications linked to these three diseases seem to be distinctly correlated with the regions where the outbreaks originated, with authors from China, for example, being much more strongly represented on SARS and Covid-19 studies, two diseases whose origin of outbreaks were attributed to this country. Middle Eastern countries, on the other hand, are exceptionally represented in the MERS literature.

The questions of where the Covid-19 literature is headed, how big it will grow in the next coming years, at what point in time the rate of publications on this topic are going to slow down (if ever) and how widely this literature is going to spread across journals and subject categories are only a few examples of questions that will be determined by time. These may also be influenced down the line by possible highly sought medical discoveries in relation to vaccine and treatment development, or lack thereof. But given the current rate at which scholarly outputs are emerging and given the extent of studies, projects, and trials that have already been conceived on this topic around the world; and also given the seemingly long-lasting and far-reaching consequences of this global emergency which have impacted on aspects of life, it will probably not be so soon before we observe a decline in Covid-19 research interest.

## References

Almeida, J., Berry, D., Cunningham, C., Hamre, D., Hofstad, M., Mallucci, L., Mcintosh, K., Tyrrell, D., 1968. Coronaviruses. Nature 220 (650), 2.

Anggraeni, S., Maulidina, A., Dewi, M.W., Rahmadianti, S., Rizky, Y.P.C., Arinalhaq, Z.F., Usdiyana, D., Nandiyanto, A.B.D., Mahdi Al-Obaidi, A.S., 2020. The deployment of drones in sending drugs and patient blood samples covid-19. Indonesian Journal of Science and Technology 5 (2).

Barmparis, G.D., Tsironis, G.P., 2020. Estimating the infection horizon of covid-19 in eight countries with a data-driven approach. Chaos, Solitons and Fractals 135.

Bekiros, S., Kouloumpou, D., 2020. Sbdiem: A new mathematical model of infectious disease dynamics. Chaos, Solitons and Fractals 136.

Boccaletti, S., Ditto, W., Mindlin, G., Atangana, A., 2020. Modeling and forecasting of epidemic spreading: The case of covid-19 and beyond. Chaos, Solitons and Fractals 135.

Bonilla-Aldana, D.K., Quintero-Rada, K., Montoya-Posada, J.P., Ramírez-Ocampo, S., Paniz-Mondolfi, A., Rabaan, A.A., Sah, R., Rodríguez-Morales, A.J., 2020. Sars-cov, mers-cov and now the 2019-novel cov: Have we investigated enough about coronaviruses?–a bibliometric analysis. Travel medicine and infectious disease 33, 101566.

Brainard, J., 2020. Scientists are drowning in covid-19 papers. Can new tools keep them afloat? Science.

Bychkova, O.V., 2020. Covid-19 and climate change reactions: Sts potential of online research. Social Anthropology.

Cavanagh, D., 2005. Coronaviridae: A review of coronaviruses and toroviruses. Coronaviruses with special emphasis on first insights concerning sars. Springer, pp. 1-54.

Chahrour, M., Assi, S., Bejjani, M., Nasrallah, A.A., Salhab, H., Fares, M., Khachfe, H.H., 2020. A bibliometric analysis of covid-19 research activity: A call for increased output. Cureus 12 (3).

Chang, L., Yan, Y., Wang, L., 2020. Coronavirus disease 2019: Coronaviruses and blood safety. Transfusion Medicine Reviews.

Chen, Y., Liu, Q., Guo, D., 2020. Emerging coronaviruses: Genome structure, replication, and pathogenesis. Journal of medical virology 92 (4), 418-423.

Cherry, J.D., Krogstad, P., 2004. Sars: The first pandemic of the 21 st century. Pediatric research 56 (1), 1-5.

Colavizza, G., Costas, R., Traag, V.A., Van Eck, N.J., Van Leeuwen, T., Waltman, L., 2020. A scientometric overview of cord-19. BioRxiv.

Cortegiani, A., Ingoglia, G., Ippolito, M., Giarratano, A., Einav, S., 2020. A systematic review on the efficacy and safety of chloroquine for the treatment of covid-19. Journal of Critical Care.

Dandekar, A., Ghai, R., 2020. Migration and reverse migration in the age of covid-19. Economic and Political Weekly 55 (19), 28-31.

De Vos, J., 2020. The effect of covid-19 and subsequent social distancing on travel behavior. Transportation Research Interdisciplinary Perspectives 5.

Dehghanbanadaki, H., Seif, F., Vahidi, Y., Razi, F., Hashemi, E., Khoshmirsafa, M., Aazami, H., 2020. Bibliometric analysis of global scientific research on coronavirus (covid-19). Medical Journal of The Islamic Republic of Iran (MJIRI) 34 (1), 354-362.

Golinelli, D., Nuzzolese, A.G., Boetto, E., Rallo, F., Greco, M., Toscano, F., Fantini, M.P., 2020. The impact of early scientific literature in response to covid-19: A scientometric perspective. medRxiv.

Haghani, M., Bliemer, M.C.J., Goerlandt, F., Li, J., 2020. The scientific literature on coronaviruses, covid-19 and its associated safety-related research dimensions: A scientometric analysis and scoping review. Safety Science 129, 104806.

Hossain, M.M., 2020. Current status of global research on novel coronavirus disease (covid-19): A bibliometric analysis and knowledge mapping. Available at SSRN 3547824.

Jarvis, C., 2020. Journals, peer reviewers cope with surge in covid-19 publications. The Scientist.

Kagan, D., Moran-Gilad, J., Fire, M., 2020. Scientometric trends for coronaviruses and other emerging viral infections. BioRxiv.

Karaseva, A., 2020. The legal void and covid-19 governance. Social Anthropology.

Knight, C.A., 1954. The chemical constitution of viruses. In: Smith, K.M., Lauffer, M.A. eds. Advances in virus research. Academic Press, pp. 153-182.

Kostoff, R.N., Morse, S.A., 2011. Structure and infrastructure of infectious agent research literature: Sars. Scientometrics 86 (1), 195-209.

Kumar, K., 2020. Author productivity of covid-19 research output globally: Testing lotka's law. Available at SSRN 3603889.

Le Bras, P., Gharavi, A., Robb, D.A., Vidal, A.F., Padilla, S., Chantler, M.J., 2020. Visualising covid-19 research. arXiv, arXiv: 2005.06380.

Lee, D., Lee, J., 2020. Testing on the move: South korea's rapid response to the covid-19 pandemic. Transportation Research Interdisciplinary Perspectives 5.

Lim, Y.X., Ng, Y.L., Tam, J.P., Liu, D.X., 2016. Human coronaviruses: A review of virus–host interactions. Diseases 4 (3), 26.

Mcintosh, K., 1974. Coronaviruses: A comparative review. Current topics in microbiology and immunology/ergebnisse der mikrobiologie und immunitätsforschung. Springer, pp. 85-129.

Mulchandani, P., 2020. Covid-19 crisis: Economic stimulus packages and environmental sustainability. Economic and Political Weekly 55 (19).

Myint, S., 1994. Human coronaviruses: A brief review. Reviews in Medical Virology 4 (1), 35-46.

Ndaïrou, F., Area, I., Nieto, J.J., Torres, D.F.M., 2020. Mathematical modeling of covid-19 transmission dynamics with a case study of wuhan. Chaos, Solitons and Fractals 135.

Olijnyk, N.V., 2015. An algorithmic historiography of the ebola research specialty: Mapping the science behind ebola. Scientometrics 105 (1), 623-643.

Postnikov, E.B., 2020. Estimation of covid-19 dynamics “on a back-of-envelope”: Does the simplest sir model provide quantitative parameters and predictions? Chaos, Solitons and Fractals 135.

Reardon, T., Mishra, A., Nuthalapati, C.S.R., Bellemare, M.F., Zilberman, D., 2020. Covid-19’s disruption of india’s transformed food supply chains. Economic and Political Weekly 55 (18), 18-22.

Ribeiro, M.H.D.M., Da Silva, R.G., Mariani, V.C., Coelho, L.D.S., 2020. Short-term forecasting covid-19 cumulative confirmed cases: Perspectives for brazil. Chaos, Solitons and Fractals 135.

Sedaju, A., Haryono, S., Anisahwati, N., 2020. Flexible work arrangement in manufacturing during the covid-19 pandemic: An evidence-based study of indonesian employees. International Journal of Advanced Science and Technology 29 (6), 3914-3924.

Sohrabi, C., Alsafi, Z., O'neill, N., Khan, M., Kerwan, A., Al-Jabir, A., Iosifidis, C., Agha, R., 2020. World health organization declares global emergency: A review of the 2019 novel coronavirus (covid-19). International Journal of Surgery 76, 71-76.

Tian, D., Zheng, T., 2015. Emerging infectious disease: Trends in the literature on sars and h7n9 influenza. Scientometrics 105 (1), 485-495.

Torres-Salinas, D., Robinson-Garcia, N., Castillo-Valdivieso, P.A., 2020. Open access and altmetrics in the pandemic age: Forescast analysis on covid-19 related literature. BioRxiv.

Van Eck, N., Waltman, L., 2010. Software survey: Vosviewer, a computer program for bibliometric mapping. scientometrics 84 (2), 523-538.

Wang, L., Wang, Y., Ye, D., Liu, Q., 2020. Review of the 2019 novel coronavirus (sars-cov-2) based on current evidence. International Journal of Antimicrobial Agents, 105948.

Wesemann, C., Pieralli, S., Fretwurst, T., Nold, J., Nelson, K., Schmelzeisen, R., Hellwig, E., Spies, B.C., 2020. 3-d printed protective equipment during covid-19 pandemic. Materials 13 (8).

Zhang, X., Ma, R., Wang, L., 2020. Predicting turning point, duration and attack rate of covid-19 outbreaks in major western countries. Chaos, Solitons and Fractals 135.

## Appendix

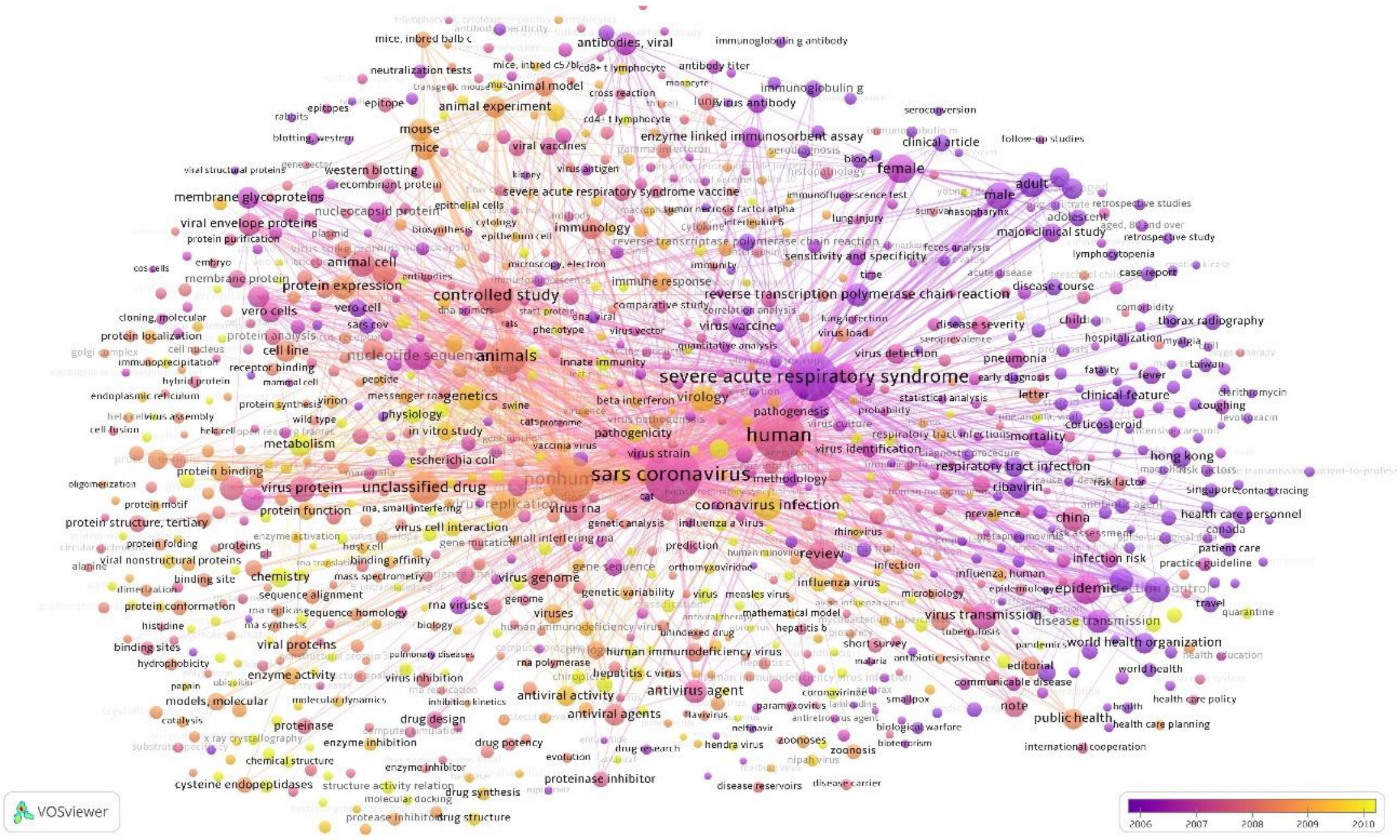

**Figure A1** The map of keyword co-occurrence for SARS overlaid with the colour-coding of the average year of publication

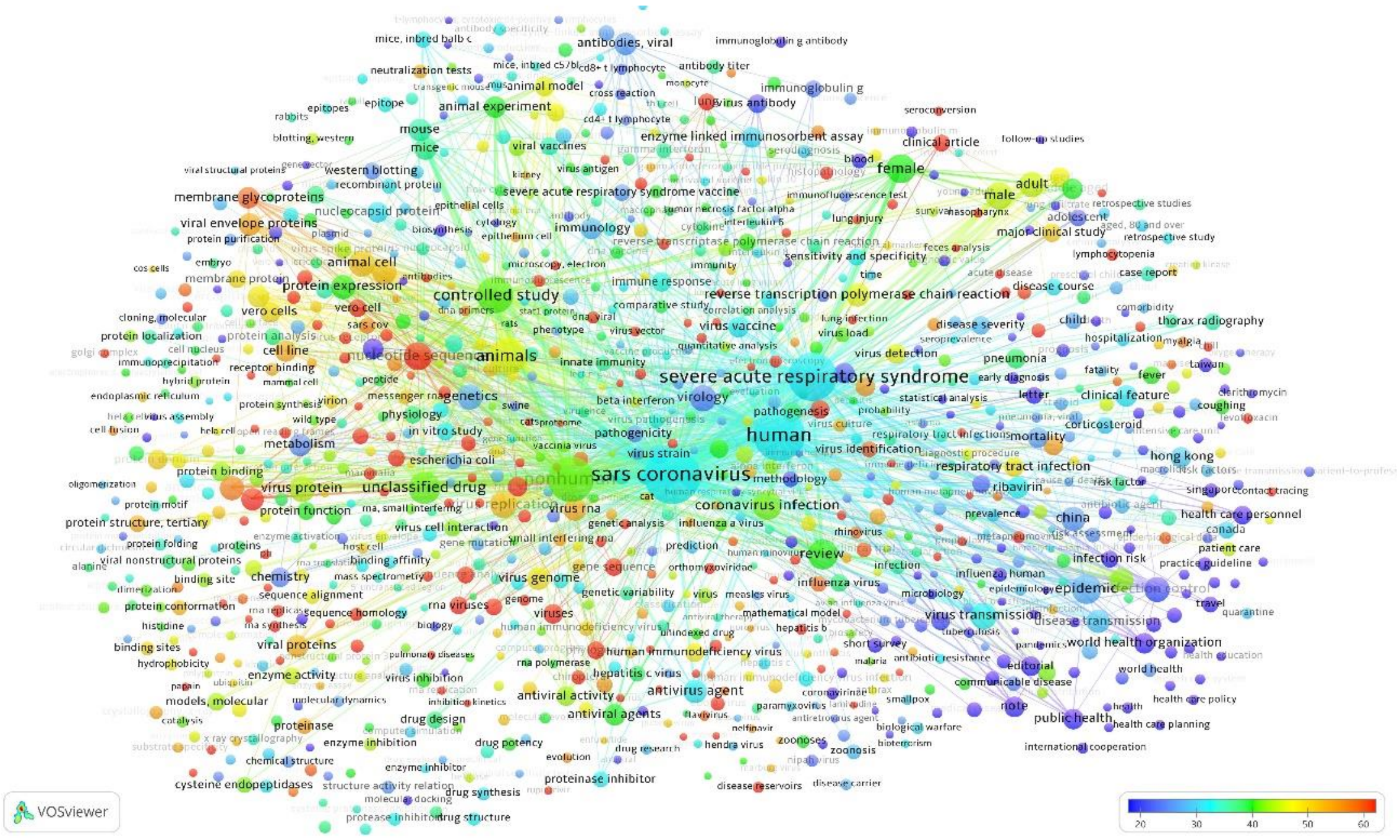

**Figure A2** The map of keyword co-occurrence for SARS overlaid with the colour-coding of the average citation number

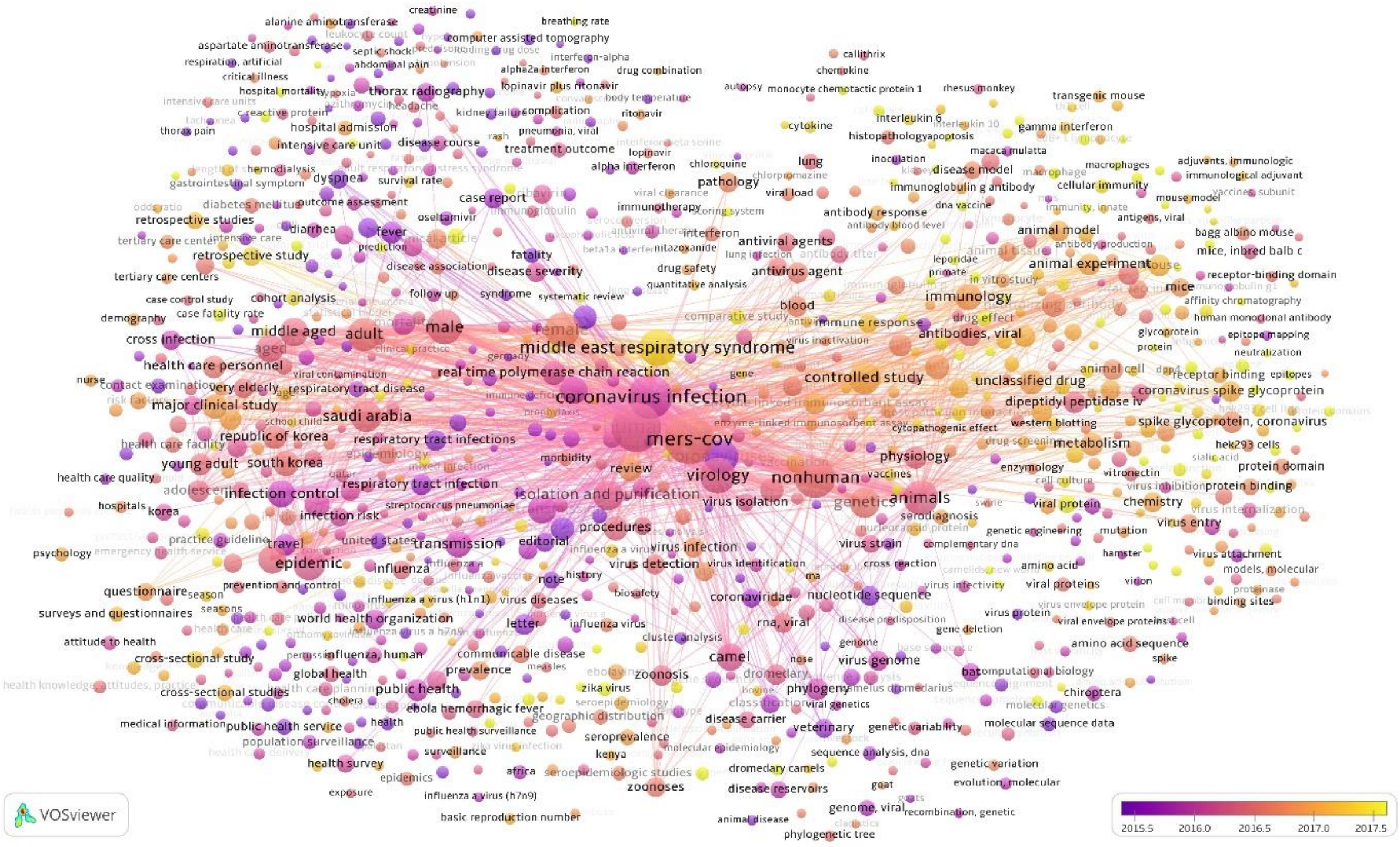

**Figure A3** The map of keyword co-occurrence for MERS overlaid with the colour-coding of the average year of publication

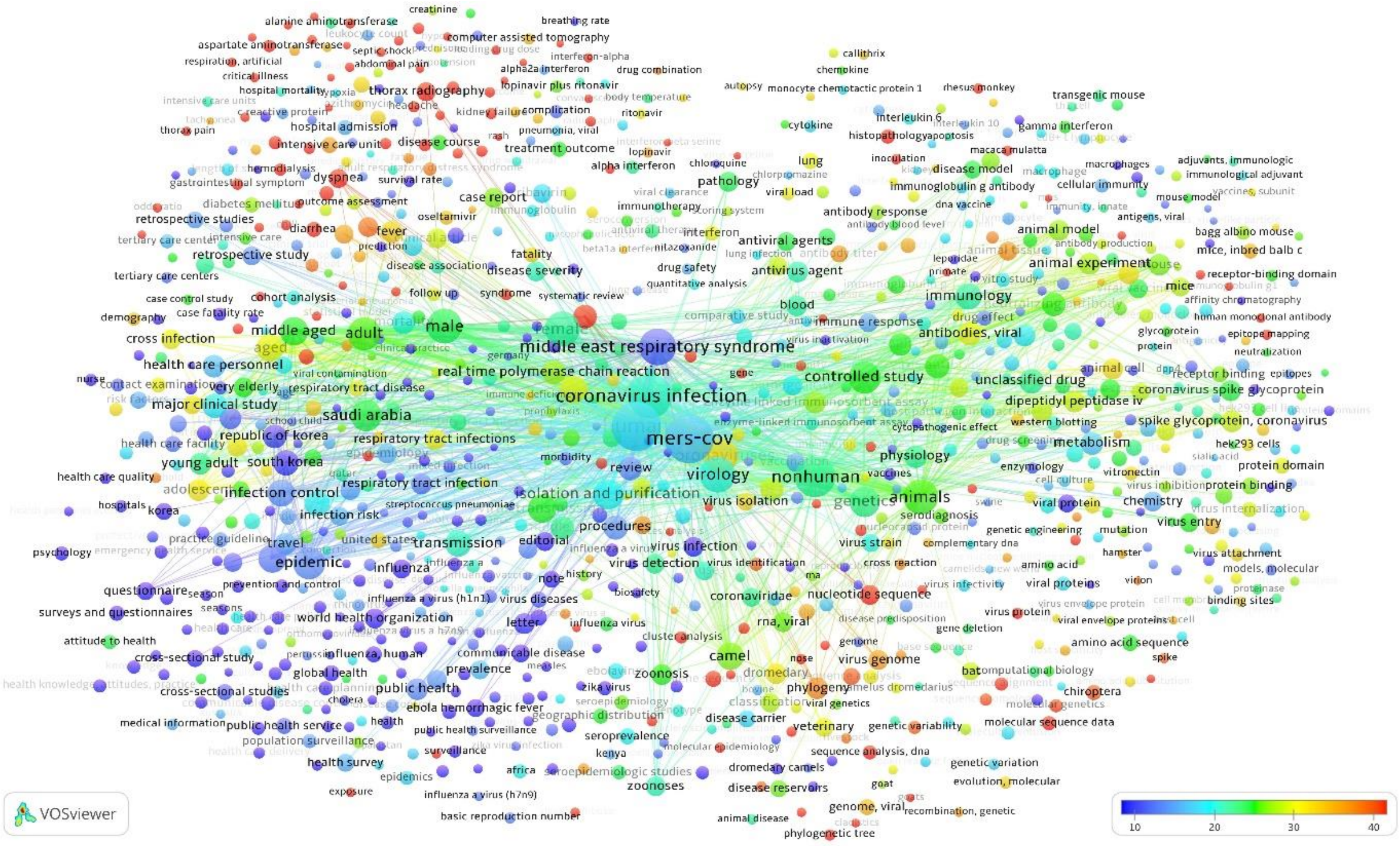

**Figure A4** The map of keyword co-occurrence for MERS overlaid with the colour-coding of the average citation number

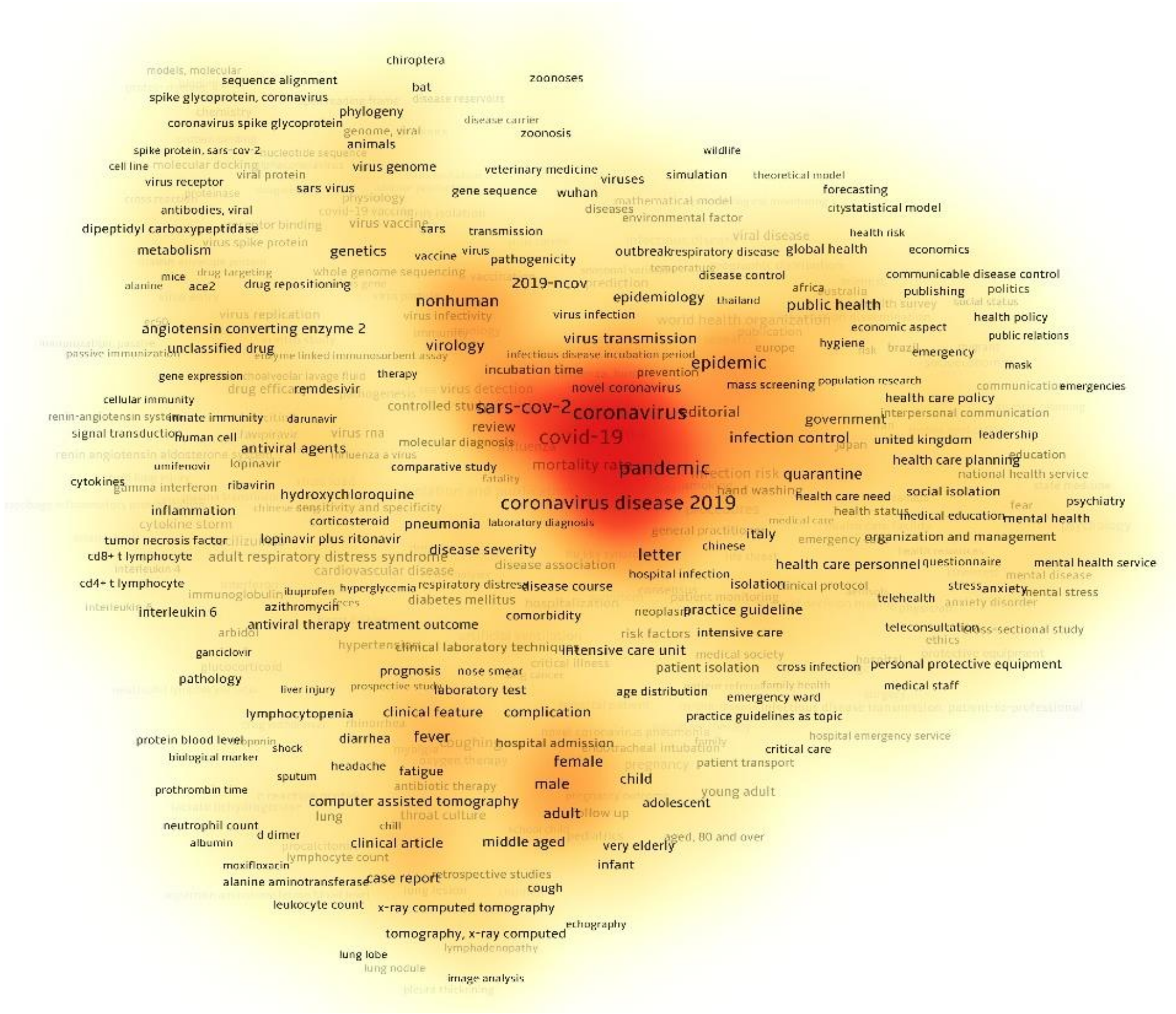

**Figure A5** The heatmap of keyword co-occurrence for Covid-19

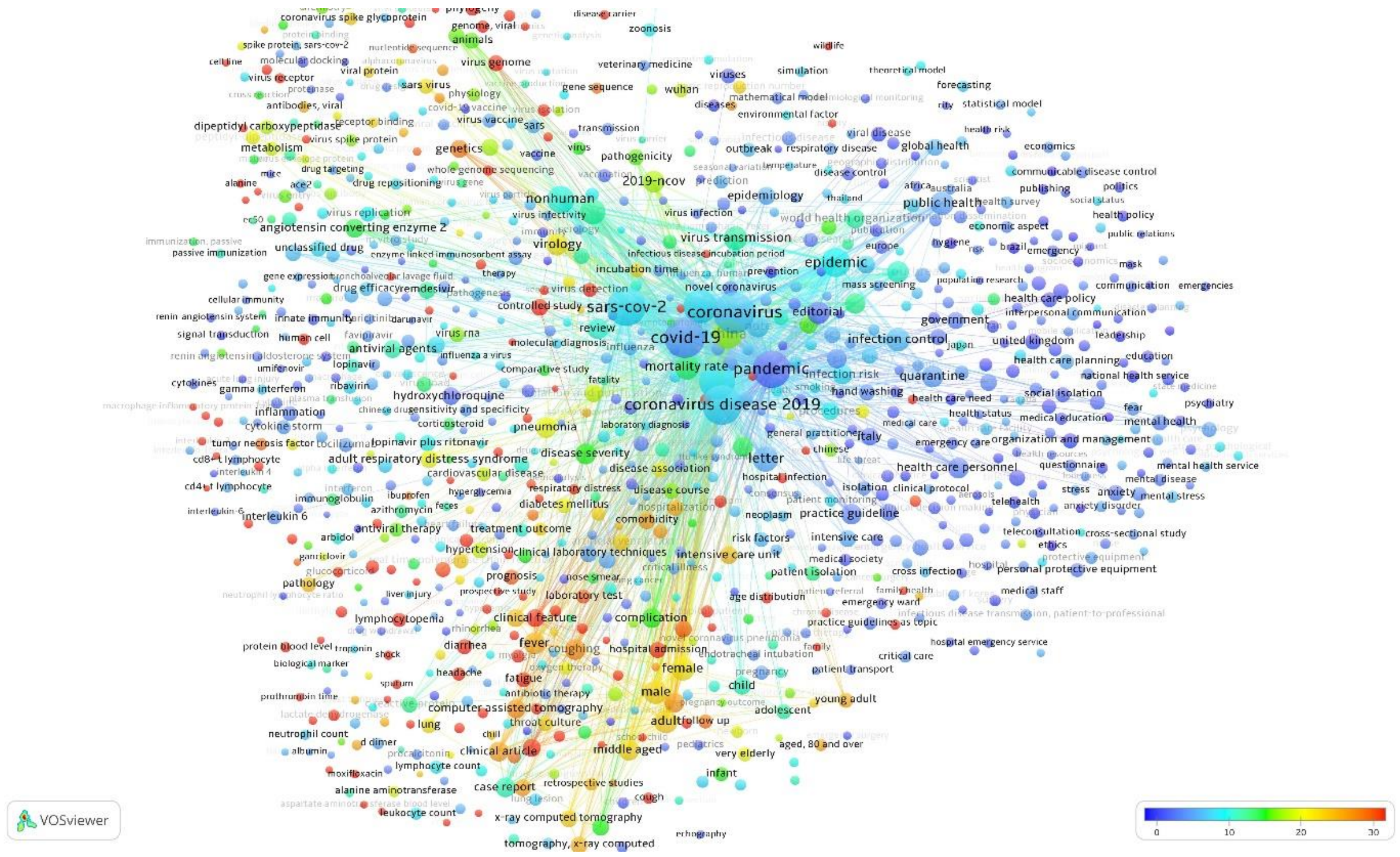

**Figure A6** The map of keyword co-occurrence for Covid-19 overlaid with the colour-coding the average citation number

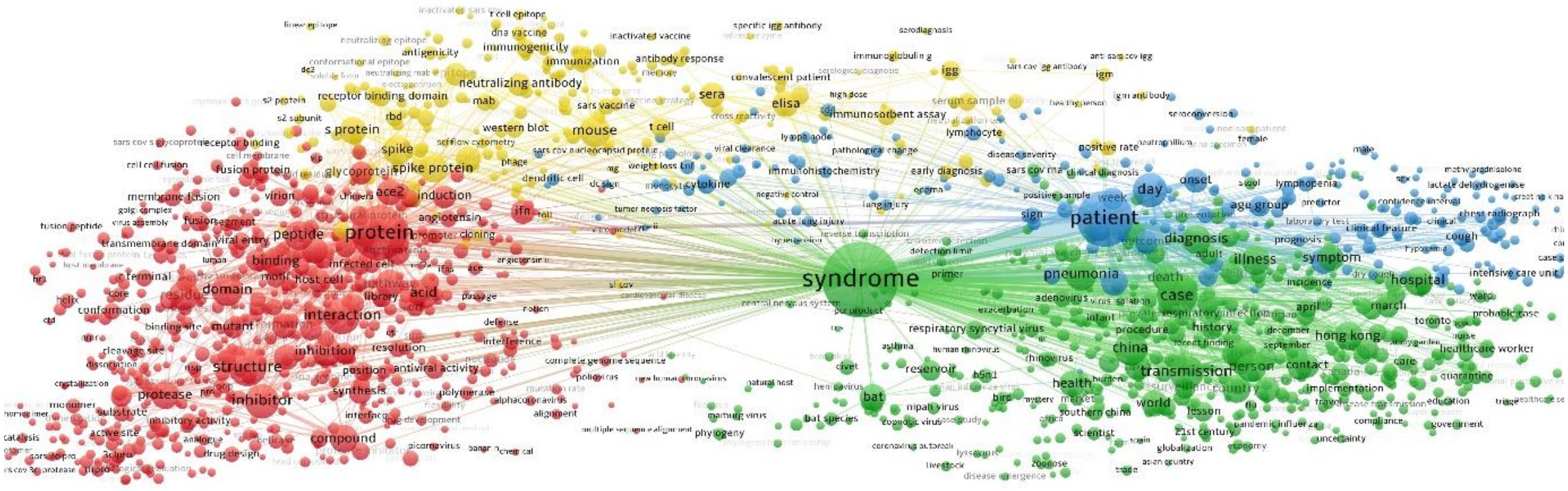

**Figure A7** The map of term co-occurrence for SARS based on text analysis of titles and abstracts

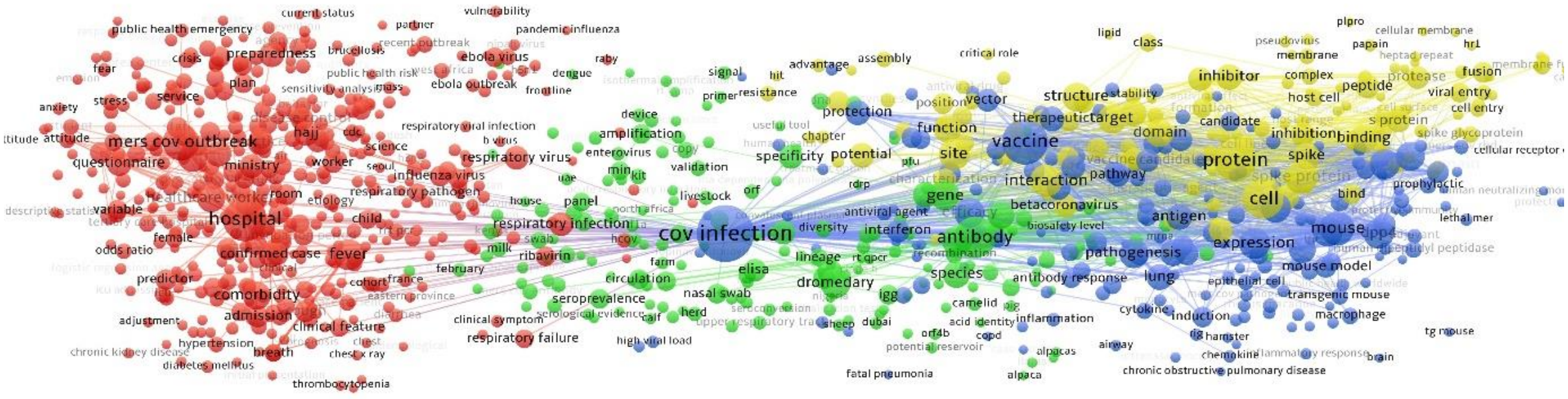

**Figure A8** The map of term co-occurrence for MERS based on text analysis of titles and abstracts

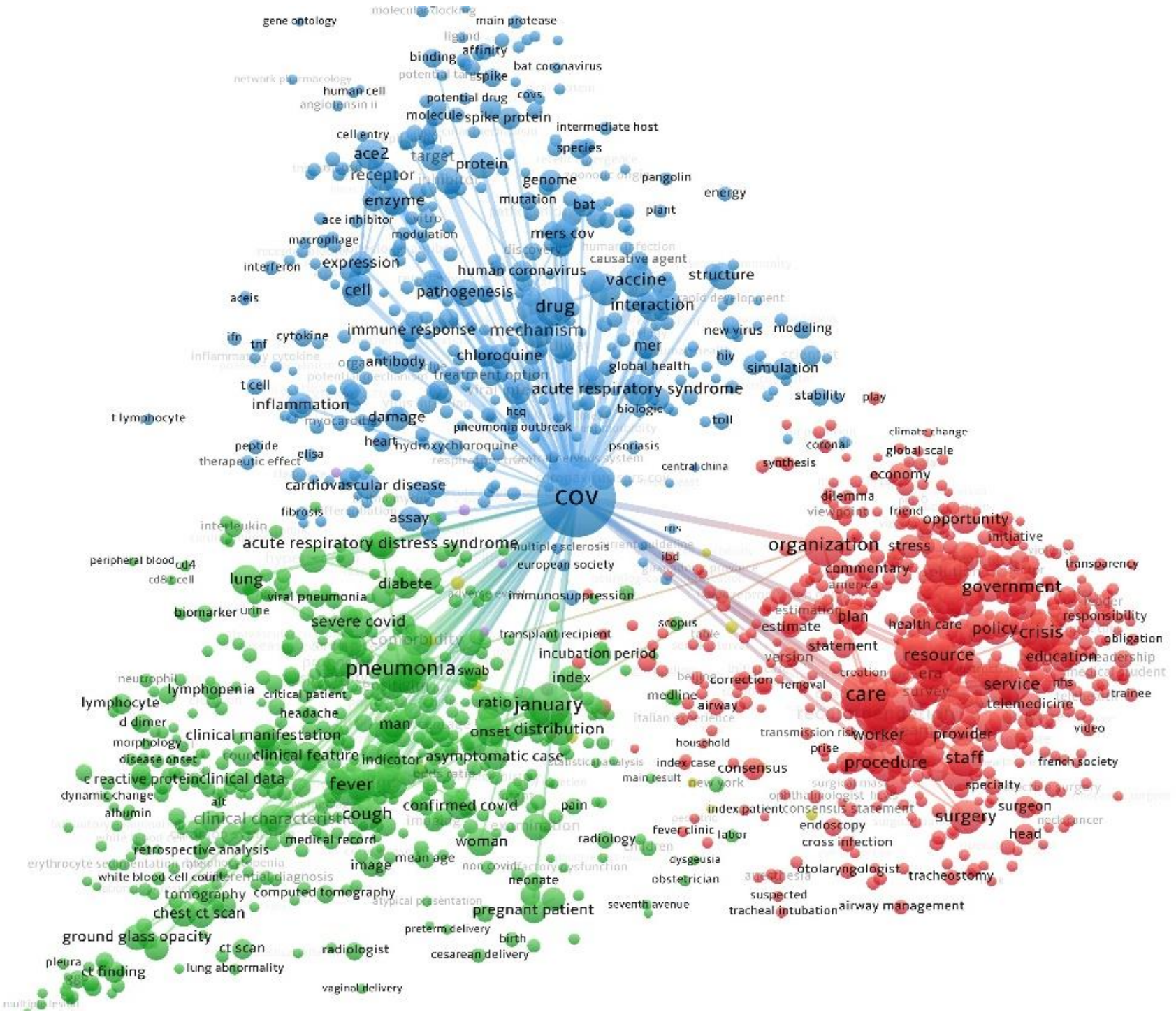

**Figure A9** The map of term co-occurrence for MERS based on text analysis of titles and abstracts

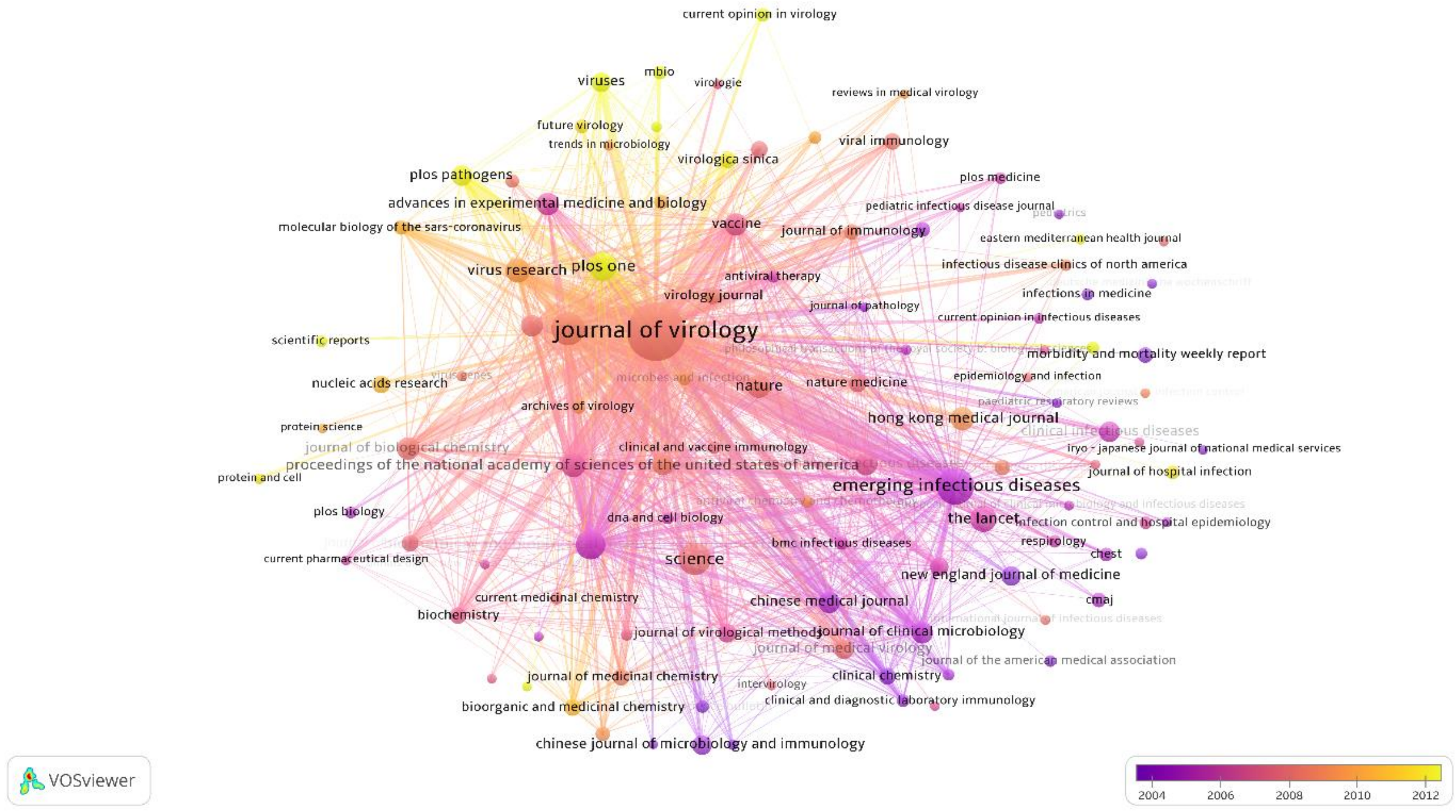

**Figure A10** The map of bibliographic coupling of sources for SARS overlaid with the colour-coding of the average publication year

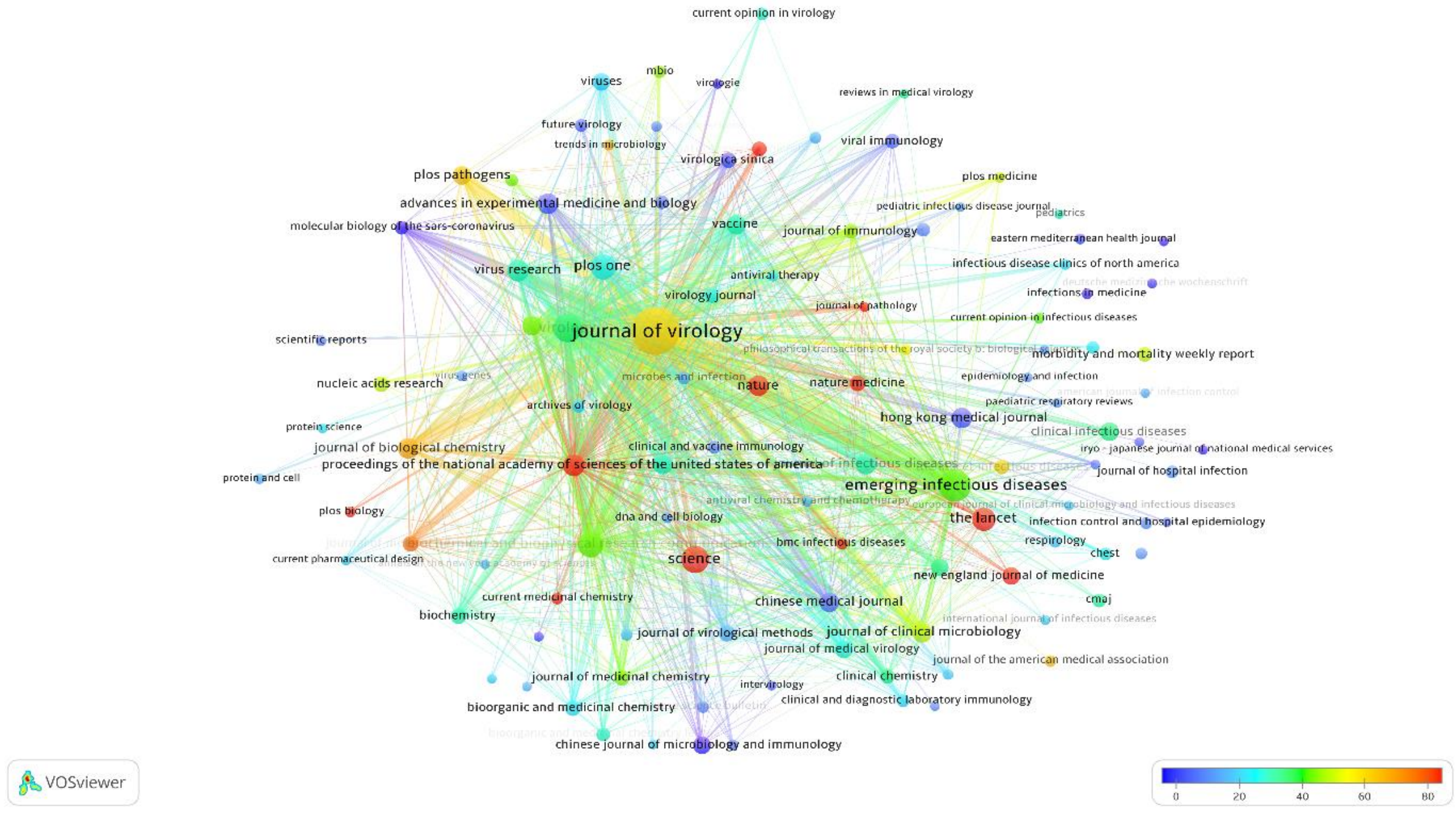

**Figure A11** The map of bibliographic coupling of sources for SARS overlaid with the colour-coding of the average citation per document

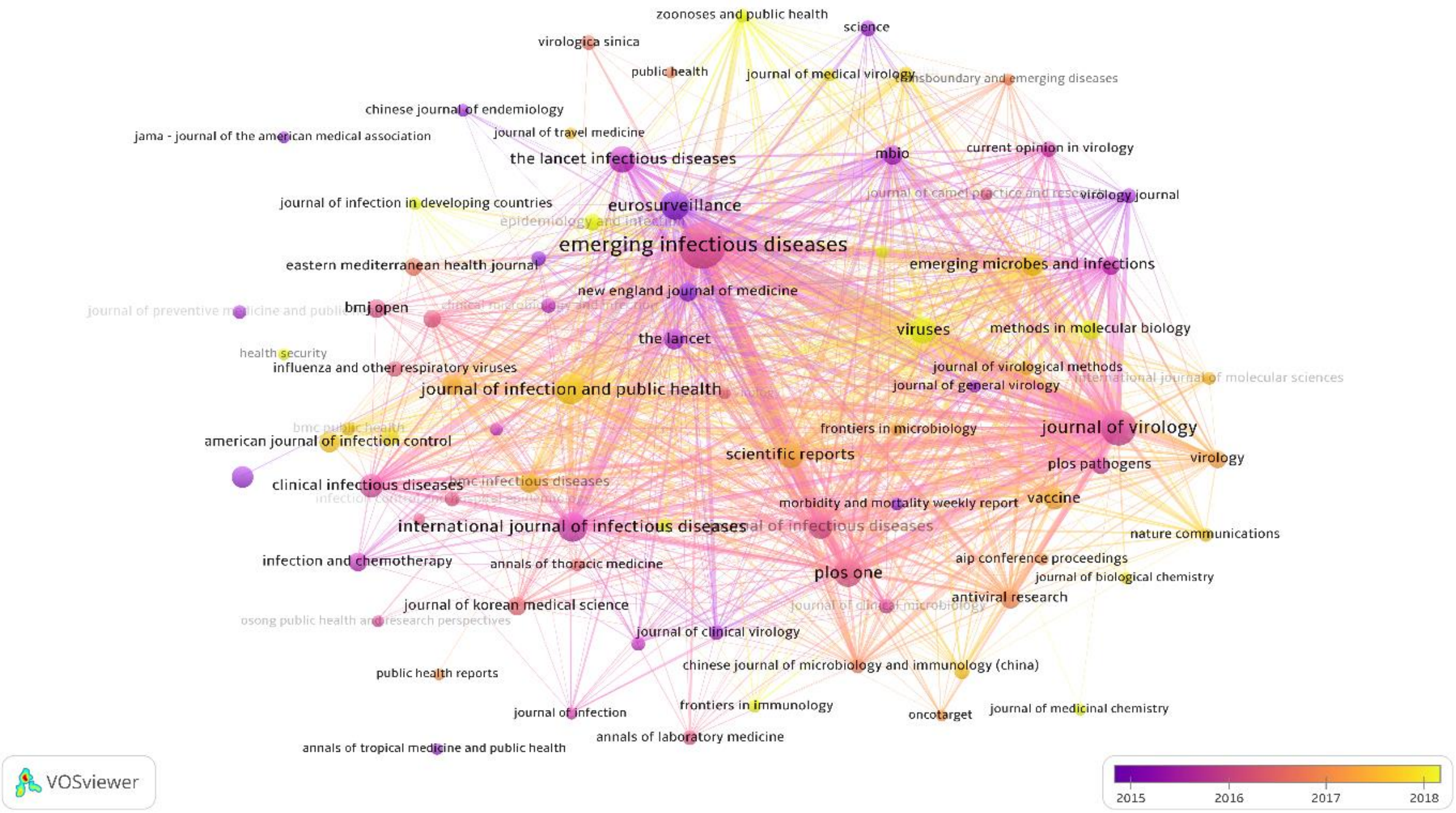

**Figure A12** The map of bibliographic coupling of sources for MERS overlaid with the colour-coding of the average publication year

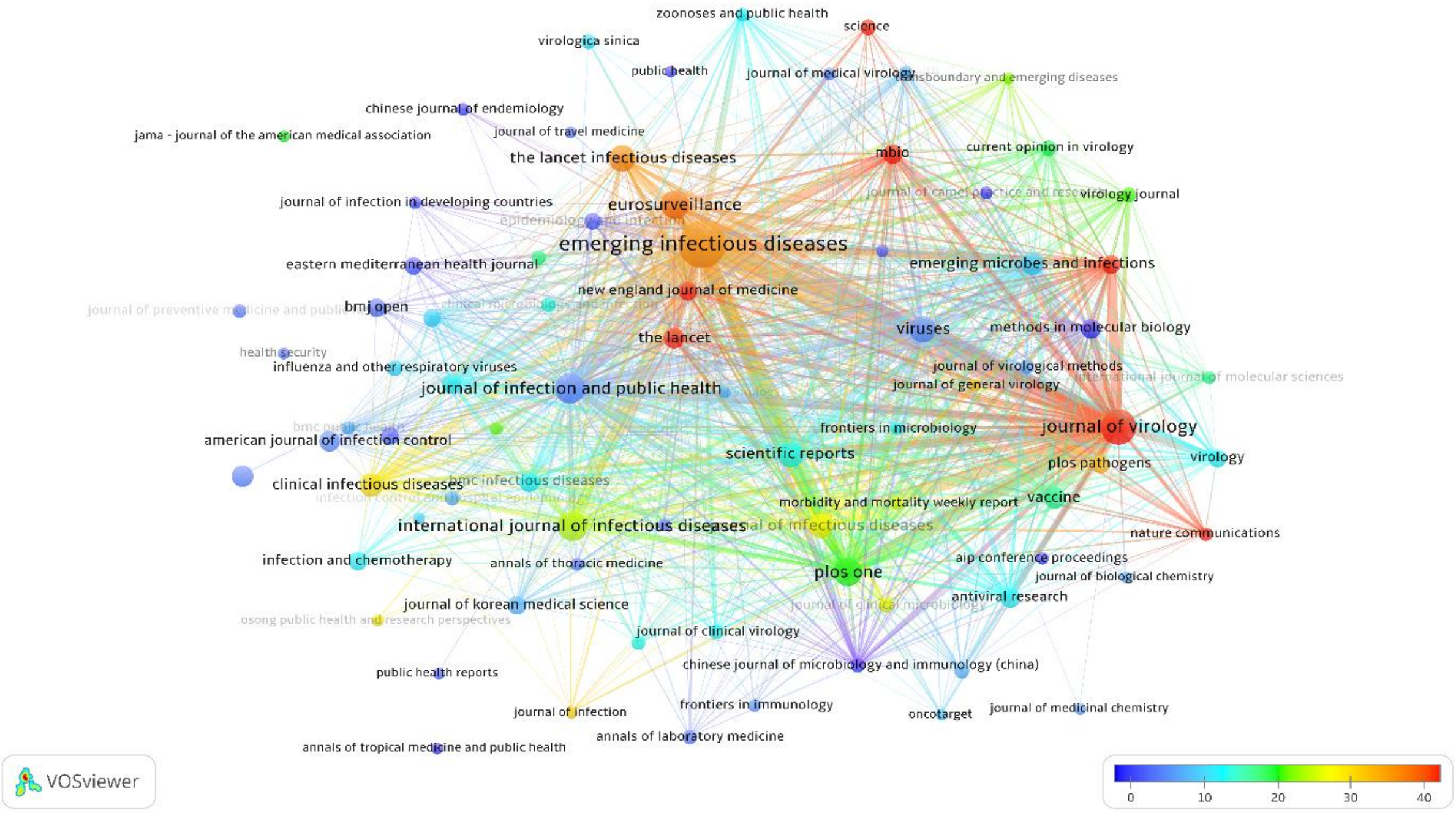

**Figure A13** The map of bibliographic coupling of sources for MERS overlaid with the colour-coding of the average citation per document

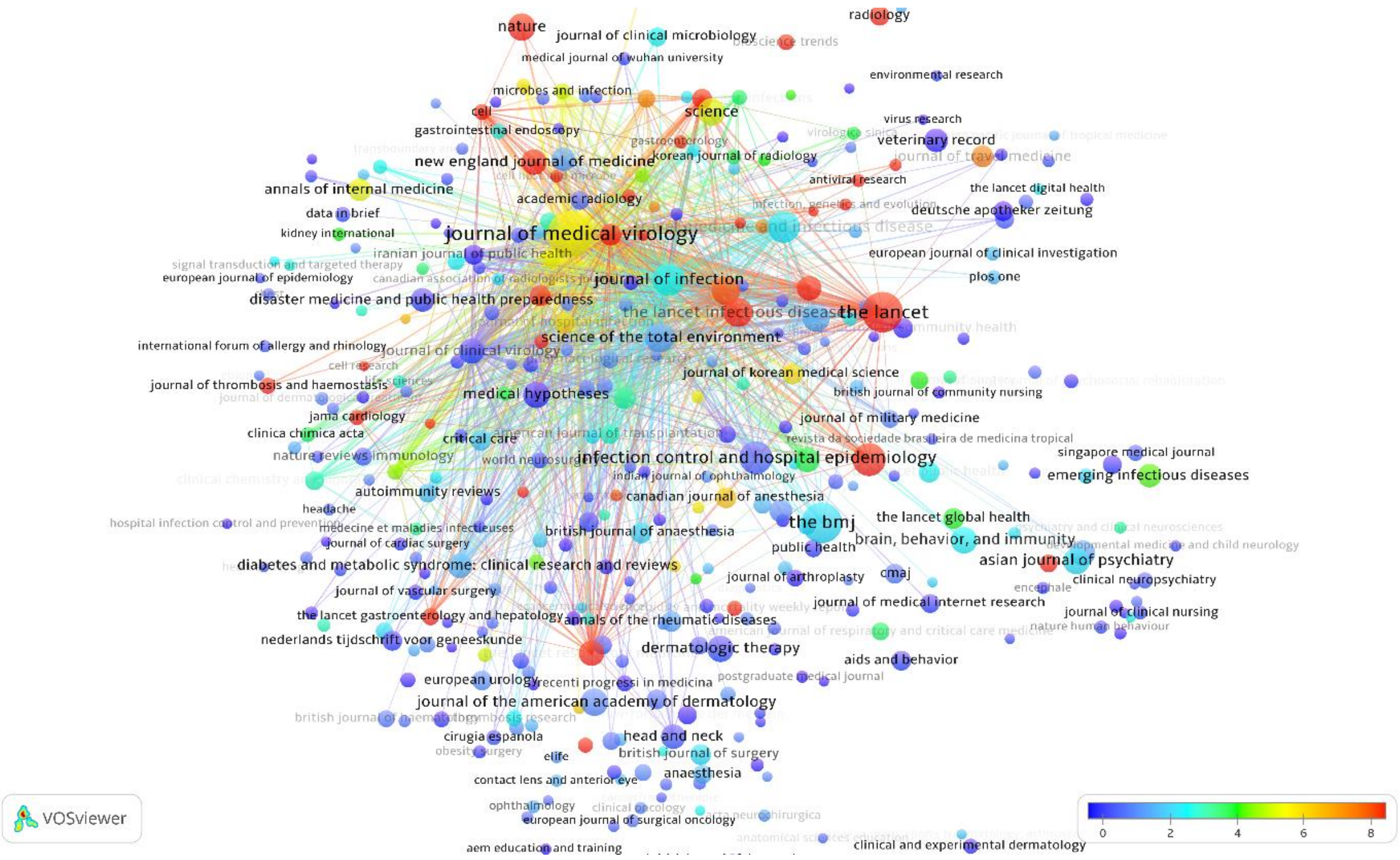

**Figure A14** The map of bibliographic coupling of sources for Covid-19 overlaid with the colour-coding of the average citation per document